\documentclass[10pt,journal,twoside]{IEEEtran}
\usepackage{booktabs} \usepackage{multirow}

\usepackage{array}
\usepackage{booktabs}

\usepackage{bm}
\usepackage{tipa}
\usepackage{dblfloatfix} %
\usepackage{mathtools}
\usepackage{algpseudocode}
\usepackage{tabularx}

\usepackage[font=small]{caption}

\usepackage{latexsym,epsf,graphicx,psfrag,amssymb,cite}
\usepackage{cite,tabularx}
\usepackage{epsf}
\usepackage{amsmath,tikz}
\usepackage{amssymb}
\usepackage{epsfig}
\usepackage{psfrag}
\usepackage{graphicx}
\usepackage{graphics}
\usepackage{multirow}
\usepackage{array}
\usepackage{enumerate}
\usepackage{enumitem} 
\usepackage{textgreek}
\usepackage{amsthm}
\usepackage[table,xcdraw]{xcolor}
\usepackage{float,lscape}
\usepackage{longtable}
\usepackage[font=small]{caption}
\usepackage{algpseudocode}
\usepackage{subcaption}
\usepackage[ruled, vlined, Algorithm]{algorithm}
\usepackage{multicol}
\usepackage{comment}
\usepackage[hyphens]{url}
\usepackage{hyperref}
\hypersetup{
    colorlinks=true,
    linkcolor=black,
    filecolor=black,      
    urlcolor=cyan,
    pdftitle={Overleaf Example},
    pdfpagemode=FullScreen,
    citecolor=cyan
    }

    \usepackage{nomencl}
\makenomenclature
\definecolor{EditorColor}{RGB}{0,0,0}          
\definecolor{ReviewerOne}{RGB}{0,0,0}        
\definecolor{ReviewerTwo}{RGB}{0,0,0}        
\definecolor{ReviewerThree}{RGB}{0,0,0}    
\definecolor{ReviewerFour}{RGB}{0,0,0}       
\definecolor{ChangeColor}{RGB}{0,0,0}       
\usepackage[normalem]{ulem}
\usepackage{balance}
\usepackage{arydshln} 
\usepackage{soul}
\usepackage[font={small}]{caption}
\usepackage{setspace}
\usepackage{mathrsfs}  
\usepackage[english]{babel}
\usepackage{scalefnt}

\allowdisplaybreaks

\newcolumntype{M}[1]{>{\centering\arraybackslash}m{#1}}
\newcolumntype{N}{@{}m{0pt}@{}}
\usepackage{nomencl}
\makenomenclature

\usepackage{colortbl}
\usepackage[table]{xcolor}
\usepackage{pifont}
\usepackage{framed}
\usepackage{soul}
\colorlet{shadecolor}{gray!15}
\usepackage[nonumberlist]{glossaries}   

\newglossarystyle{ieeeacronyms}{%
    {\glslistinit\begin{LaTeXdescription}}{\end{LaTeXdescription}}%
  \renewcommand*{\glsgroupheading}[1]{}%
}
\setglossarystyle{ieeeacronyms}

\renewcommand{\arraystretch}{1.3}

\algrenewcommand\algorithmicrequire{\textbf {Input:}}
\algrenewcommand\algorithmicensure{\textbf{Output:}}
\title{\textcolor{black}{Performance Analysis of MR-NOMA with Arbitrary Number of Users and Symbol Rates}}

\author{Zainab Khader, \IEEEmembership{Graduate Student Member, IEEE}, Arafat Al-Dweik \IEEEmembership{Senior Member, IEEE}, and  Emad Alsusa \IEEEmembership{Senior Member, IEEE}
\thanks{Zainab Khader is a graduate student with the Department of Computer and Information Engineering, Khalifa University, Abu Dhabi, P.O.Box 127788, UAE (e-mail: 100061989@ku.ac.ae). }
\thanks{Arafat Al-Dweik is with the 6G Research Center and Department of Computer and Information Engineering,
Khalifa University, Abu Dhabi, P.O.Box 127788, UAE. (e-mail: arafat.dweik@ku.ac.ae, dweik@fulbrightmail.org).}
\thanks{Emad Alsusa is with the Department of Electrical and Electronic Engineering, The University of Manchester, Manchester, M13 9PL, UK. (e-mail: e.alsusa@manchester.ac.uk).}
}
\makeglossaries
\newacronym{3gpp}{3GPP}{3rd generation partnership project}
\newacronym{6g}{$6$G}{Sixth generation}
\newacronym{iot}{IoT}{Internet of Things}
\newacronym{ack}{ACK}{acknowledgment}
\newacronym{scs}{SCS}{sub-carrier spacing}
\newacronym{v2v}{V2V}{vehicle to vehicle}
\newacronym{acm}{ACM}{adaptive coding and modulation}
\newacronym{arq}{ARQ}{automatic repeat request}
\newacronym{harq}{HARQ}{hybrid-ARQ}
\newacronym{per}{PER}{packet error rate}
\newacronym{b5g}{5G+}{beyond 5G}
\newacronym{pdf}{PDF}{probability density function}
\newacronym{ber}{BER}{bit error rate}
\newacronym{bc}{BC}{Broadcast}
\newacronym{spc}{SPC}{super-position coding}
\newacronym{vr}{VR}{virtual reality}
\newacronym{ar}{AR}{augmented reality}
\newacronym{bler}{BLER}{bLock error rate}
\newacronym{bpsk}{BPSK}{Binary phase shift keying}
\newacronym{btfdps}{BTFDPS}{block time-frequency domain packet scheduling}
\newacronym{itu}{ITU}{international telecommunication union}
\newacronym{ptd}{PTD}{packet time delay}
\newacronym{pdr}{PDR}{packet drop rate}
\newacronym{cqi}{CQI}{channel quality indicator}
\newacronym{crc}{CRC}{cyclic redundancy check}
\newacronym{iui}{IUI}{inter-user interference}
\newacronym{ngma}{NGMA}{next generation multiple access}
 
\newacronym{csi}{CSI}{channel state information}
\newacronym{voip}{VoIP}{voice  over  internet  protocol} 
\newacronym{msg3}{MSG3}{message-3} 
\newacronym{3g}{$3$G}{third generation} 
\newacronym{4g}{$4$G}{fourth generation} 
\newacronym{5g}{$5$G}{Fifth generation} 
\newacronym{wifi}{WiFi}{wireless fidelity}
\newacronym{fec}{FEC}{forward error correction}
\newacronym{mimo}{MIMO}{multiple-input multiple-output}
\newacronym{mmimo}{mMIMO}{massive multiple-input multiple-output}
\newacronym{htc}{HTC}{human-type communication}
\newacronym{d2d}{D2D}{device-to-device}
\newacronym{cr}{CR}{cognitive radio}
\newacronym{icr}{ICR}{interweave \gls{cr}}
\newacronym{ucr}{UCR}{underlay \gls{cr}}
\newacronym{bb}{BB}{branch and bound}
\newacronym{lc bb}{LC BB}{least cost branch and bound}
\newacronym{mmwave}{mm-Wave}{millimeter wave}
\newacronym{wpans}{WPANs}{wireless personal area networks}
\newacronym{nb}{NB}{narrow band}
\newacronym{di}{DI}{deferred iterations}
\newacronym{dl}{DL}{downlink}
\newacronym{ldpc}{LDPC}{low density parity check}
\newacronym{tpc}{TPC}{turbo product code}

\newacronym{es}{ES}{early stop}
\newacronym{ed}{ED}{error detection}

\newacronym{gop}{GoP}{group of pictures}
\newacronym{hevc}{HEVC}{high efficiency video coding}

\newacronym{ici}{ICI}{inter-carrier interference}

\newacronym{lte}{LTE}{long-term evolution}

\newacronym{mac}{MAC}{medium access control}
\newacronym{mib}{MIB}{management information base}

\newacronym{mpeg}{MPEG}{moving picture experts group}
\newacronym{mse}{MSE}{mean squared error}

\newacronym{nack}{NACK}{negative acknowledgment}
\newacronym{ofdm}{OFDM}{orthogonal frequency-division multiplexing}
\newacronym{noma}{NOMA}{non-orthogonal multiple access}

\newacronym{oma}{OMA}{orthogonal multiple access}

\newacronym{pec}{PEC}{parity error checking}

\newacronym{phy}{PHY}{physical layer}
\newacronym{prb}{PRB}{physical resource block}
\newacronym{psnr}{PSNR}{peak signal to noise ratio}

\newacronym{qam}{QAM}{quadrature amplitude modulation}
\newacronym{qos}{QoS}{quality-of-service}
\newacronym{qp}{QP}{quantization parameter}
\newacronym{qpsk}{QPSK}{quadrature phase shift keying}
\newacronym{ini}{INI}{inter-numerology-interferance}
\newacronym{rlc}{RLC}{radio link control}

\newacronym{ser}{SER}{symbol error rate}
\newacronym{sic}{SIC}{successive interference cancellation}
\newacronym{snr}{SNR}{signal-to-noise ratio}
\newacronym{zf}{ZF}{zero-forcing}
\newacronym{rsma}{RSMA}{rate splitting multiple access}
\newacronym{egs}{EGS}{exhaustive grid search}
\newacronym{pa}{PA}{power allocation}
\newacronym{bis}{BIS}{bisection search}
\newacronym{sr-harq}{SR-HARQ}{selective repeat \gls{harq}}
\newacronym{ma}{MA}{multiple access}
\newacronym{tdps}{TDPS}{time domain packet scheduling}
\newacronym{tpcs}{TPCs}{turbo product codes}
\newacronym{ul}{UL}{uplink}
\newacronym{ue}{UE}{user equipment}
\newacronym{umts}{UMTS}{universal mobile telecommunications system}
\newacronym{tsc}{TSC}{Time Sensitive Communication}
\newacronym{urllc}{URLLC}{ultra reliable low latency communication}
\newacronym{hiho}{HIHO}{hard-input hard-output}
\newacronym{siso}{SISO}{single-input single-output}
\newacronym{bch}{BCH}{Bose–Chaudhuri–Hocquenghem}
\newacronym{lfsr}{LFSR}{linear  feedback  shift register}
\newacronym{std}{STD}{standard deviation}
\newacronym{llr}{LLR}{log-likelihood ratio}
\newacronym{fm}{$F_M$}{feedback message}
\newacronym{rtt}{RTT}{round trip time}
\newacronym{sw}{SW}{stop-and-wait}
\newacronym{srnoma tm}{SR-TM}{single-symbol rate NOMA transmission mode}
\newacronym{srnoma}{SR-NOMA}{single-symbol rate NOMA}
\newacronym{oma tm}{OMA-TM}{OMA transmission mode}
\newacronym{gbn}{GBN}{go-back-N}
\newacronym{sr}{SR}{single-symbol rate}
\newacronym{awgn}{AWGN}{additive white Gaussian noise}
\newacronym{mpsk}{MPSK}{M-ary phase shift keying}
\newacronym{milp}{MILP}{Mixed integer linear programming}
\newacronym{iid}{i.i.d.}{independent and identically distributed}
\newacronym{mrc}{MRC}{maximum ratio combining}
\newacronym{hdd}{HDD}{hard decision decoding}
\newacronym{sdd}{SDD}{soft decision decoding}
\newacronym{fer}{FER}{frame error rate}
\newacronym{bps}{bps}{bits per second}
\newacronym{pps}{pps}{packets per second}
\newacronym{fps}{fps}{frames per second}
\newacronym{tti}{TTI}{transmission time interval}
\newacronym{sinr}{SINR}{signal-to-interference plus noise ratio}
\newacronym{ami}{AMI}{accumulated mutual information}
\newacronym{tnr}{TNR}{transmission number relaying}
\newacronym{bs}{BS}{base station}
\newacronym{ee}{EE}{energy efficiency}

\newacronym{gsm}{GSM}{global system for mobile communications}
\newacronym{cdma}{CDMA}{code division multiple access}
\newacronym{wcdma}{W-CDMA}{wideband-CDMA}
\newacronym{fdd}{FDD}{frequency division duplexing}
\newacronym{td}{TD}{time division}  
\newacronym{scdma}{SCDMA}{synchronous CDMA}
\newacronym{tdd}{TDD}{time division duplexing}
\newacronym{hsdpa}{HSDPA}{high speed downlink packet access}
\newacronym{hsupa}{HSUPA}{high speed uplink packet access}
\newacronym{hspa+}{HSPA+}{evolved high speed packet access}
\newacronym{e-utra}{E-UTRA}{evolved universal terrestrial radio access}
\newacronym{eutran}{EUTRAN}{evolved universal terrestrial radio access network}
\newacronym{um}{UM}{unacknowledged mode}
\newacronym{am}{AM}{acknowledged mode}
\newacronym{sn}{SN}{sequence number}
\newacronym{pdu}{PDU}{protocol data unit}
\newacronym{amd}{AMD}{acknowledged mode data}
\newacronym{rrc}{RRC}{radio resources control}
\newacronym{ir}{IR}{incremental redundancy}
\newacronym{imt2020}{IMT-$2020$}{International Mobile Telecommunications-$2020$}

\newacronym{wimax}{WiMAX}{worldwide  interoperability  for  microwave access}
\newacronym{psdu}{PSDU}{physical data service unit}
\newacronym{spid}{SPID}{subpacket identifier}
\newacronym{nr}{NR}{new radio}
\newacronym{cbg}{CBG}{codeblock-group}
\newacronym{tb}{TB}{transport block}
\newacronym{wlan}{WLAN}{wireless local area network}
\newacronym{back}{BAck}{block \gls{ack}}
\newacronym{snmp}{SNMP}{simple network management protocol}
\newacronym{sta}{STA}{wireless station}
\newacronym{ssrc}{SSRC}{station short retry count}
\newacronym{slrc}{SLRC}{station long retry count}
\newacronym{ac}{AC}{access category}
\newacronym{tc}{TC}{traffic category}
\newacronym{tms}{TMs}{transmission modes}
\newacronym{tm}{TM}{transmission mode}
\newacronym{wpan}{WPAN}{wireless personal area network}
\newacronym{lr}{LR}{low-rate}
\newacronym{mm}{mm}{millimeter}
\newacronym{hr}{HR}{high-rate}
\newacronym{mpdu}{MPDU}{\gls{mac} protocol data unit}
\newacronym{mcs}{MCS}{modulation and coding scheme}
\newacronym{mtc}{MTC}{machine type communication}
\newacronym{mmtc}{mMTC}{massive machine type communication}
\newacronym{enb}{eNB}{evolved node B}
\newacronym{blc}{BLC}{bit level combining}
\newacronym{slc}{SLC}{symbol level combining}
\newacronym{aoi}{AoI}{age of information}
\newacronym{sbs}{SBS}{small base station}
\newacronym{mbs}{MBS}{macro base station}
\newacronym{wap}{WAP}{wireless access point}
\newacronym{tcp}{TCP}{transmission control protocol}
\newacronym{cran}{CRAN}{cloud-radio access network}
\newacronym{bpu}{BPU}{baseband processing unit}
\newacronym{rrh}{RRH}{remote  radio  head}
\newacronym{rau}{RAU}{radio access unit}
\newacronym{rl}{RL}{reinforcement learning}
\newacronym{wsn}{WSN}{wireless sensor network}
\newacronym{dcf}{DCF}{decode-and-forward}
\newacronym{af}{AF}{amplify-and-forward}
\newacronym{df}{DF}{demodulated-and-forward}
\newacronym{swipt}{SWIPT}{simultaneous wireless information and power transmission}
\newacronym{rf}{RF}{radio frequency}
\newacronym{uav}{UAV}{unmanned aerial vehicle}
\newacronym{ai}{AI}{artificial intelligence}
\newacronym{nom}{NOM}{non-orthogonal multiplexing}

\newacronym{eharq}{E-HARQ}{early HARQ}
\newacronym{ap}{AP}{access point}
\newacronym{cpc}{CPC}{complementary punctured codes}
\newacronym{lp}{LP}{low priority}
\newacronym{hp}{HP}{high priority}

\newacronym{lpn}{LPN}{\gls{lp}  node}
\newacronym{hpn}{HPN}{\gls{hp}  node}

\newacronym{pu}{PU}{primary user}
\newacronym{su}{SU}{secondary user}

\newacronym{fpa}{FPA}{fixed \gls{pa}}
\newacronym{apa}{APA}{adaptive \gls{pa}}
\newacronym{map}{MAP}{maximum a priori}
\newacronym{ml}{ML}{maximum likelihood}
\newacronym{cc}{CC}{control center}
\newacronym{be}{BE}{best-effort}
\newacronym{pd}{PD}{power domain}
\newacronym{embb}{eMBB+}{enhanced mobile broadband}
\newacronym{urllc+}{URLLC+}{ultra-reliable low-latency communications}
\newacronym{mui}{MUI}{multi-user interference}
\newacronym{mrnoma tm}{MR-TM}{multi-symbol rate NOMA transmission mode}
\newacronym{noma tm}{NOMA TM}{NOMA transmission mode}
\newacronym{pnoma}{PL-NOMA}{partial NOMA}
\newacronym{fb}{FB}{frequency band}
\newacronym{fbs}{FBs}{frequency bands}
\newacronym{fdma}{FDMA}{frequency division multiple access}
\newacronym{tdma}{TDMA}{time division multiple access}
\newacronym{op}{OP}{outage probability}
\newacronym{mld}{MLD}{maximum likelihood detector}
\newacronym{jmumld}{JMuMLD}{joint multiuser MLD}

\newacronym{sir}{SIR}{signal-to-interference ratio}
\newacronym{sjd}{SJD}{successive joint decoding}
\newacronym{ic}{IC}{interference cancellation}
\newacronym{jml}{JML}{joint maximum likelihood}

\newacronym{vlc}{VLC}{visible light communications}
\newacronym{sdma}{SDMA}{space division multiple access}
\newacronym{must}{MUST}{multi-user superposition transmission}
\newacronym{ldm}{LDM}{layered division multiplexing}

\newacronym{bea}{BEA}{band elimination algorithm}
\newacronym{mr}{MR}{multi-rate}

\newacronym{cdf}{CDF}{cumlatived distribution function}

\definecolor{EditorColor}{RGB}{0,0,0}          
\definecolor{ReviewerOne}{RGB}{0,0,0}        
\definecolor{ReviewerTwo}{RGB}{0,0,0}        
\definecolor{RTwoColor}{RGB}{0,0,0}

\newcommand{\RTwo}[1]{\textcolor{RTwoColor}{#1}}

\definecolor{ReviewerThree}{RGB}{0,0,0}    
\definecolor{ReviewerFour}{RGB}{0,0,0}       
\definecolor{ChangeColor}{RGB}{0,0,0}  
\newcommand{\Rone}[1]{{\color{ReviewerOne}#1}}
\newcommand{\Rtwo}[1]{{\color{ReviewerTwo}#1}}
\newcommand{\Rthree}[1]{{\color{ReviewerThree}#1}}
\newcommand{\Rfour}[1]{{\color{ReviewerFour}#1}}
\newif\ifcolorfulbody
\colorfulbodyfalse

\newcounter{rvcomment}



\begin{document}
\scalefont{0.97}
\maketitle
\begin{abstract}
 
This paper analytically investigates the \gls{ber} performance of downlink
\gls{mr}-\gls{noma} over Nakagami-$\frak{m}$ fading channels for an arbitrary
number of users and symbol rates. The analysis is first developed for a three-user case,
then generalized to systems with more than three users, yielding an exact \gls{ber}
expression for the first detected user and an accurate closed-form
approximations for the remaining users under the perfect-\gls{sic}
assumption. These results are further unified into a general average-\gls{ber}
expression over Nakagami-$\frak m$ fading for arbitrary user numbers and symbol-rate configurations. A
symbol-pattern-dependent \gls{sinr} analysis further shows how the
symbol-duration diversity enables inherent interference cancellation, explaining the \gls{ber} advantage of \gls{mr}-\gls{noma} over conventional
\gls{sr}-\gls{noma}. An average \gls{ber} minimization problem is also formulated and numerically solved to optimize the users' power allocation, with simulation results corroborating the analytical findings. Overall, \gls{mr}-\gls{noma} is shown to offer greater flexibility in user pairing while maintaining reliable \gls{ber} performance and accommodating diverse \gls{qos} requirements.
\end{abstract}
\glsresetall
\glsresetall
\section{Introduction}\label{sec:introduction}
According to GSMA \cite{gsma2024mobile}, global mobile internet connectivity surpassed 4.7 billion users by the end of 2024, marking an increase of 2.2 billion since 2015. Moreover, by 2030, \gls{6g} technology is expected to emerge, extending beyond \gls{5g} capabilities and enabling unprecedented advances in latency, speed, reliability, and device intelligence \cite{wang2023road,zhang20196g,gustavsson2021implementation}. This evolution will require accommodating the massive connectivity of wireless devices, coupled with highly heterogeneous \gls{qos} demands, posing a critical challenge that will define the design of future wireless networks \cite{10597056}\cite{tataria20216g}. 

Meeting this challenge calls for advancing \gls{ma} techniques beyond classical communication-based approaches toward the innovative \gls{ngma} framework \cite{liu2022evolution,liu2022developing}. In particular, \gls{ngma} aims to support massive connectivity in \gls{6g} and accommodate diverse use cases enabled by widespread adoption of \gls{iot} technologies, thereby enabling better optimized operations compared to existing \gls{ma} schemes \cite{10562043}.

As a prominent member of the \gls{ngma} family, \gls{noma} has been recognized as a key candidate for \gls{6g} networks
\cite{liu2022developing}. Since it was shown that \gls{oma} techniques achieve a strictly smaller capacity region than non-orthogonal transmission, particularly in \gls{bc} channels, and following the introduction of \gls{spc} and \gls{sic} as an effective means of achieving the capacity region of degraded \gls{bc} channels
\cite{cover1972broadcast, cover1981achievable, molavianjazi2015second,
jorswieck2024next}, attention has shifted to non-orthogonal transmission approaches
\cite{7676258, ding2017application, mao2022rate, mu2023noma,
ding2024next, RSMA}. The main principle of
\gls{noma} is to allow multiple users to share the same transmission
resources simultaneously, while employing \gls{sic} at the receiver to separate the users' signals. Although other detection schemes can also be used, \gls{sic} is generally preferred for its lower complexity. In this work, we consider \gls{pd}-\gls{noma}, in which each user is assigned
a distinct power level.

The primary objective of \gls{ngma} is to support extremely high user connectivity while simultaneously accommodating a diverse range of use cases. Such diversity is driven by the varying \gls{qos} requirements of emerging \gls{6g} applications, spanning data rate, reliability, and latency. To this end, \gls{ngma} must evolve into a \gls{qos}-aware framework. \gls{noma} offers a solid foundation in this regard; however, it still requires further development to fully realize this vision \cite{wang2023road}. \Rtwo{\label{R_2_4_1} For instance, \gls{mr}-\gls{noma} introduced in \cite{UL_MRNOMA_Arafat}, and the focus of this paper, extends the traditional \gls{noma} design by allowing users to transmit with different symbol rates, making it more suitable for supporting various \gls{qos} requirements.}

\subsection{Related Work}\label{section: Related Work}
\Gls{5g} wireless communication networks have shifted the data-centric approach of legacy mobile generations toward a service-based wireless system that supports diverse use-cases with heterogeneous requirements \cite{popovski20185g}. The prospective use cases have been categorized into a few major service groups; mainly \gls{embb}, \gls{urllc}, and \gls{mmtc} \cite{popovski20185g}. Moreover, to embrace the diversity in service requirements, a standardized mixed numerology is adopted in \gls{5g} \gls{nr}. \Rfour{\label{R_4_3} The mixed numerology is a \gls{3gpp} terminology that refers mainly to the choice of subcarrier spacing and the resulting \gls{ofdm} time-frequency structure \cite{Numerology1}, \cite{Numerolog2}}. Mixed numerology was introduced for a flexible radio access technology \cite{3gpp_ts_38_211_v18_6_0}, which can be supported
by time domain or frequency domain multiplexing. For example, low-latency use cases can use shorter symbol duration while delay-insensitive ones, such as \gls{iot} devices for smart city and smart factory \cite{7829438}, employ narrower spacings and longer symbol durations, all within the same carrier \cite{3gpp_ts_38_211_v18_6_0}. In contrast, \gls{urllc} applications, such as \gls{v2v} communications, internet services, and industrial automation \cite{korrai2020joint}, might require numerologies with wide \gls{scs}, consequently, a higher symbol rate to fulfill stringent latency requirements \cite{GuardBand2}.

As such, recent research focuses on optimizing systems under mixed numerology, such as addressing \gls{ini}, which is an interference introduced due to the non-orthogonality among subcarriers, \cite{INI1}, \cite{INI2}, guard-band and windowing design strategies to mitigate and reduce \gls{ini} effect \cite{guardBand1, GuardBand2}, and resource scheduling \cite{ResourceAllocation}. Moreover, \gls{ma} and multi-user transmission strategies in the presence of mixed numerology have also been studied recently \cite{Mcwade1, Mcwade2, MIMONumerology}. In \cite{Mcwade1}, the achievable rates of two numerologies in the \gls{ul} were analyzed using the conventional \gls{noma} technique. Building on \cite{Mcwade1}, \cite{Mcwade2} proposed a resource allocation scheme that maximizes the sum rate while ensuring the minimum rate requirements of individual users. In \cite{MIMONumerology}, the authors considered \gls{mmimo} \gls{dl} systems and employed subcarrier-wise zero-forcing pre-coders to mitigate intra-numerology interference. 

More recently, \Rfour{a notable line of work is a single-carrier \gls{mr}-\gls{noma}, which treats mixed-numerology in terms of symbol-rate heterogeneity in the time domain \cite{UL_MRNOMA_Arafat}, \cite{yahya2024designMR_NOMA}, \cite{SumRateMax_MR_NOMA}}. Specifically, \cite{UL_MRNOMA_Arafat} introduced the concept of \gls{mr}-\gls{noma}, where users with different symbol rates can be paired, thereby exploiting the relation between symbol rate and symbol energy as an additional degree of freedom. The study demonstrates that such flexibility enhances system design by increasing the number of supported users or improving throughput. At the same time, optimal power selection enables both high- and low-rate users to achieve lower \glspl{ber} and improved transmission reliability. Further, \cite{yahya2024designMR_NOMA} analyzed the \gls{dl} of \gls{mr}-\gls{noma} and derived closed-form \gls{ber} expressions for the two-user case. Moreover, in \cite{SumRateMax_MR_NOMA}, the authors derived the average \gls{ber} for the \gls{dl} \gls{mr}-\gls{noma} over a Nakagami-$\frak{m}$ channel in wireless sensor networks, and maximized the sum-rate by optimizing the users' pairing and rate requirements.


\subsection{Motivation and Contributions}\label{section:Motivation}
Although various aspects of \gls{sr}-\gls{noma} have been extensively investigated in the literature, \gls{mr}-\gls{noma} is a relatively new concept, and associated research remains scarce. Based on an extensive literature survey, and to the best of the authors’ knowledge, the \gls{ber} performance of \gls{mr}-\gls{noma} with arbitrary symbol durations and its generalization to an arbitrary number of users have not been previously investigated, and the corresponding power-coefficient optimization problem has not been previously studied.  \Rtwo{\Rfour{}\label{R_2_2}Another important motivation stems from the heterogeneous rate and reliability requirements of emerging wireless networks. Unlike \gls{sr}-\gls{noma}, \gls{mr}-\gls{noma} offers an additional degree of design flexibility by allowing users to employ different symbol durations and, consequently, different transmission rates. However, the flexibility also results in a more intricate interference structure, making the selection of system parameters, particularly the power coefficients, considerably more challenging. Therefore, a general analytical characterization of the \gls{ber} is essential not only for evaluating the reliability of \gls{mr}-\gls{noma}, but also for exploiting its additional degrees of freedom through efficient system optimization.
Motivated by the potential of \gls{mr}-\gls{noma} to enhance network performance, this work provides a comprehensive analysis under conditional and average \gls{ber} constraints. Furthermore, increasing the number of users would increase transmitter/receiver complexity, and thus, it is essential to optimize the system design.} \Rfour{\label{R_4_2}The main contributions are summarized as follows:
\begin{itemize}
\item Derives an exact \gls{ber} for the first and second detected users
in the 3-user case $(N=3)$ with arbitrary symbol rates, as well as a
closed-form \gls{ber} expression for the third detected user with
$\boldsymbol T=[4,2,1]$.\label{R_4_6} This case
serves as the first nontrivial extension beyond the previously reported
two-user \gls{mr}-\gls{noma} case in~\cite{yahya2024designMR_NOMA},
allowing the derivation pattern required for the arbitrary user
generalization to be identified.
\item Identifies and exploits the recurring analytical pattern in the \gls{ber} expressions of $N=3$ to generalize the analysis to $N>3$, yielding an exact expression for $U_1$ and accurate closed-form approximations for the remaining users under the perfect-\gls{sic} assumption. These results are subsequently unified into a general closed-form average-\gls{ber} expression over Nakagami-$\tilde{m}$ fading for arbitrary user numbers and symbol-rate configurations.
\item Provides an analytical comparison between \gls{mr}-\gls{noma} and conventional \gls{sr}-\gls{noma}, revealing how symbol-duration diversity enables coherent combining and inherent interference cancellation, thereby improving \gls{ber} performance and reducing \gls{sic} error propagation. 
\item Quantifies the \gls{ber} improvement achievable through optimized power allocation in \gls{mr}-\gls{noma} using an average-\gls{ber} minimization framework. The study reveals how symbol-energy diversity, and the resulting increase in effective signal combinations, jointly enhance system reliability and reduce the users' average \gls{ber}.
\item Extensive analytical and simulation results are presented
to assess the performance of the proposed scheme under
various conditions. The results confirmed the superiority of \gls{mr}-\gls{noma} over \gls{sr}-\gls{noma} in terms of \gls{ber}.
\item The analytical expressions and Monte Carlo simulation framework are made available as MATLAB code in \cite{Paper_Code}.
\end{itemize}}

\Rtwo{\subsection{Notation  \label{R_2_3}}\label{Sec:notation}}
The notations used throughout the article are as follows. Upper case bold letters such as $\mathbf{M}$ denote matrices, while row and column vectors are denoted by italic upper and lower case bold letters such as $\boldsymbol{M}$ and $\mathbf{m}$. Sets are defined as $\mathbb{I}=\{I_1,I_2,\dots\}$, where $\mathbb{R}$, $\mathbb{C}$, and $\mathbb{Z}$ denote the sets of real, complex, and integer numbers, respectively. 
 The Euclidean norm is defined as $\Vert{\cdot}\Vert$ and $[.]^\top$ is the transpose, $\Pr(.)$ is the probability of an event. 
 The \gls{pdf} and \gls{cdf} are denoted by $f(.)$ and $F(.)$, respectively. 
 The standard Gaussian $Q$-function is given by $Q(\cdot)$ and $\binom{n}{k}$ represents the Binomial coefficients, 
  $\lceil \cdot \rceil$ is the ceiling operator, and $\lfloor \cdot \rfloor$ is the floor operator. The subscript $n$ indicates that the variable belongs to user $n$ and the tilde denotes the trial values.


\subsection{Paper Organization}\label{section:Paper Organization}
The remainder of this paper is organized as follows. Sec.~\ref{sec:System Model} introduces the system and channel models. Sec.~\ref{sec:BER Analysis} presents the conditional \gls{ber} analysis for the three-user case. Sec.~\ref{sec:BER Analysis_N>3} extends the conditional \gls{ber} analysis to the case of more than three users. Sec.~\ref{Sec:Average BER} derives the average \gls{ber} over Nakagami-$\frak{m}$ fading. Sec.~\ref{sec:Power Optimization} introduces the power optimization algorithm. Sec.~\ref{sec:Numerical Results} discusses the numerical results, highlighting key insights into system performance. Finally, Sec.~\ref{section:Conlusion} concludes the article and outlines future work.

\section{System and channel Models}\label{sec:System Model}
\subsection{Transmitted Signal Model}
\Rtwo{\label{R_2_5}This work considers a \gls{siso} downlink \gls{mr}-\gls{noma} where the \gls{bs} serves $N$ users, $U_1,U_2,\dots, U_N$. The users' distances from the \gls{bs} are denoted as $\mathbf{d}=[d_1,d_2,\dots,d_N]$ where $d_i \geq d_j~ \forall i\geq j$. The \gls{bs} allocates $U_n$ a power coefficient $\alpha_n$, and the set of all power coefficients is denoted as $\boldsymbol{\alpha}=[\alpha_1, \alpha_2,\dots,\alpha_N]$, $\sum_n \alpha_n=1$. In this work, we consider that power is allocated based on the statistical \gls{csi}. Thus, the power is allocated such that $\alpha_i\geq \alpha_j~ \forall i \geq j$, which is a necessary condition to provide reliable performance and enable fairness among users \cite{DistanceNOMA}}. The normalized symbol duration of user $n$ is denoted by $T_n$, $n \in \{1,2,\dots,N\}$, \Rthree{}\label{R_3_4}$T_n \in \mathbb{Z^+}$, and  $\boldsymbol{T} = [T_1, T_2, \dots, T_N]$ denotes the symbol durations vector. \Rfour{\label{R_4_4}\label{R_4_5} In practical systems, the values of $T_n$ are
the effective symbol durations selected according to the users' \gls{qos}
requirements. Users with delay-sensitive or high-rate services may use
short symbol durations, whereas delay-tolerant or reliability-oriented users may use long symbol durations to benefit from a larger
accumulated symbol energy.} Without loss of generality, the users are indexed such that $T_1>T_2>\dots>T_N$. Therefore, $U_1$ has the lowest symbol rate, is the farthest from the \gls{bs}, and is assigned the highest power coefficient. \Rthree{\label{R_3_2}Further, $
\frac{T_{1}}{T_n}\triangleq \delta_n$ indicates the number of transmitted symbols by $U_n$ within $T_1$. For analytical tractability, and without loss of generality, the minimum symbol duration $T_N$ is normalized to unity. Thus, the $j$th symbol of $U_n$ occupies a time interval of,
\begin{equation}
(j-1)T_n+1 \leq \ell \leq jT_n,
\qquad j=1,\ldots,\delta_n.
\label{eq: sample interval}
\end{equation}
and its entries are
\begin{equation}
\underbrace{
s_{n,j}\!\left[(j-1)T_n+1\right],\,
s_{n,j}\!\left[(j-1)T_n+2\right],\,
\ldots,\,
s_{n,j}\!\left[jT_n\right]
}_{T_n\text{ samples}}
\label{eq:s_n_j}
\end{equation}
\Rtwo{\label{R_2_4_2}\Rthree{\label{R_3_5}}Accordingly, all users' data symbols in $T_1$ period can be represented as $\mathbf{S}_N=[\mathbf{s}_1^{\top};\mathbf{s}_2^{\top};\ldots;
\mathbf{s}_N^{\top}]\in\mathbb{C}^{N\times\delta_N}$, which can be expressed as shown in \eqref{transmit signal matrix}
\begin{figure*}
\Rthree{
\begin{equation}
\mathbf{S}_{N}=\left[ 
\begin{array}{c}
\underbrace{s_{1,1}[1]\text{, }s_{1,1}[2]\text{, }\ldots \text{, }%
s_{1,1}[\delta _{N}]}_{T_{1}} \\ 
\underbrace{s_{2,1}[1],\,\ldots ,\,s_{2,1}[T_{2}]}_{T_{2}}\text{, }%
\underbrace{s_{2,2}[T_{2}+1],\,\ldots ,\,s_{2,2}[2T_{2}]}_{T_{2}}\text{, }%
\dots \text{, }\underbrace{s_{2,\delta _{2}}[(\delta
_{2}-1)T_{2}+1],\,\ldots ,\,s_{2,\delta _{2}}[\delta _{N}]}_{T_{2}} \\ 
\vdots  \\ 
s_{N,1}[1]\text{, }s_{N,2}[2]\text{, }\dots \text{ , }s_{N,\delta
_{N}}[\delta _{N}]%
\end{array}%
\right].
\label{transmit signal matrix}
\end{equation}}
--------------------------------------------------------------------------------------------------------------------------------------------------------
\end{figure*}}
Moreover, it can be concluded from \eqref{transmit signal matrix} that $
s_{n,j}[\ell]=s_{n,\left\lceil \frac{\ell}{T_n} \right\rceil}[\ell]
=s_{n,\left\lceil \frac{\ell \delta_n}{\delta_N} \right\rceil}[\ell],
$ where $\left\lceil \cdot\right\rceil$ is the ceiling operator.}

The data symbols $s_{n,j}$ are selected uniformly from a \gls{bpsk} constellation with a unity minimum distance. 
The transmitted \gls{mr}-\gls{noma} signal can be expressed as $\mathbf{x} = \begin{bmatrix} x[1], x[2], \dots, x[\delta_N] \end{bmatrix}^\top = \sum_n \mathbf{x}_n$, 
where $\mathbf{x}_n = \sqrt{\alpha_n P_{\mathrm{t}}} \mathbf{s}_n$, $\alpha_n$ is the power coefficient of $U_n$. The coefficients are assigned such that $\alpha_1 > \alpha_2 > \dots > \alpha_N$, ensuring reliable detection at the users’ receivers, $\sum_i \alpha_i=1$. Moreover, $P_{\mathrm{t}}$ is the maximum transmit power, which is considered unity throughout this work.
Consequently, during a $T_{1}$ window, $U_n$ would have $\delta_n$ different data symbols, and thus, 
\begin{equation}\label{eq:x_ell-1}
x[\ell ]=\sum\nolimits_{n=1}^{N}\sqrt{\alpha _{n}}s_{n,\left\lceil \frac{\ell }{
T_{n}}\right\rceil }[\ell ].
\end{equation}
For notational simplicity, the second subscript is dropped in the rest of this work; thus, $s_n[\ell]$ is used. 
\subsection{Received Signal Model}
By assuming quasi-static fading, and a uniform sampling period $T_N$ across all receivers, the received sequence at $U_n$ can be written as,
\begin{equation}
\label{eq:Received Signal model}
    \mathbf{r}_n = h_n \mathbf{x} + \mathbf{w}_n
\end{equation}
 where $\mathbf{r}_n=[r_n[1],r_n[2],\dots,r_n[\delta_N]]^{\top}$, $h_n=\frac{\grave{h}_n}{\sqrt{d_n^\eta}}$, $\grave{h}_n$ and $d_n$ are the channel gain and the distance between the \gls{bs} and \( U_n \), respectively, 
$\mathbf{\mathrm{w}}_n \in \mathbb{C}^{\delta_N\times 1}$, where $w_{n,\ell}\sim \mathcal{CN}(0, \sigma^2_{\mathrm{w},n})$ are \gls{iid} \gls{awgn}, and $\sigma^2_{\mathrm{w},n}$ is $U_n$ receiver noise power. Typically, $\sigma^2_{\mathrm{w},n}$ is equal for all users, and thus it can be expressed as $\sigma^2_{\mathrm{w}}$.
\subsection{Receiver Design}
\color{black}
Conditioned on $h_n$ and $\mathbf{S}$, the received samples are \gls{iid} complex Gaussian random variables. Consequently, the \gls{mr}-\gls{noma} \gls{jmumld} can be expressed as,
\begin{equation}
\{\hat{{s}}_{1},\hat{\mathbf{s}}_{2},\dots ,\hat{\mathbf{s}}_{N}\}=\arg
\min_{\tilde{\mathbb{S}}}\Vert \mathbf{r}_{n}-h_{n}\tilde{\mathbf{x}}\Vert_{2}^{2}
\label{jmld}
\end{equation}%
where $\tilde{\mathbb{S}}=\{{\tilde{s}}_{1},{\tilde{\mathbf{s}}}_{2},\dots,{\tilde{\mathbf{s}}}_{N}\}$, $ \{\tilde{\mathbf{s}}_n, \hat{\mathbf{s}}_n\}\in \mathbb{C}^{\delta_n\times 1}$, are respectively the trial and estimated data symbols for all users. 
Similar to \gls{sr}-\gls{noma}, the  \gls{jmumld} complexity is high \cite{assaf2019exact}, particularly for a large number of users. Therefore, the \gls{sic} receiver is considered next. It is worth noting that in \gls{dl}-\gls{noma}, the \gls{sic} and \gls{jmumld} detectors offer equivalent \gls{ber} for both \gls{sr} and \gls{mr}-\gls{noma} \cite{assaf2019exact, yahya2024designMR_NOMA}.

\subsection{\Gls{sic} Receiver}
 Using \gls{sic}, the symbols of $U_1$, which carry the largest power factor, are first detected, where the detector is equivalent to single-user \gls{mld}. By noting that the received signal of $U_1$ is oversampled, the \gls{mld} can be expressed as
\begin{align}
\hat{s}_{1} &=\arg \min_{\tilde{{s}}_{1}}\Vert {\mathbf{r}}%
_{1}-h_{1}\sqrt{\alpha _{1}}\tilde{{s}}_{1}\Vert _{2}^{2}  \notag \\
&=\arg \min_{\tilde{{s}}_{1}}\text{ }\sum_{\ell =1}^{T_{1}}\left\vert {r}_{1}[\ell]-h_{1}{\sqrt{\alpha _{1}}}\tilde{s}%
_{1}\right\vert^{2}.
\label{eq:coherant Detector}
\end{align}
\Rthree{\label{R_3_7}Since $h_1$ and $s_1$ remain constant over the $T_1$ samples, and terms independent of the trial symbol do not affect the \gls{ml} decision, the detection metric can equivalently be expressed as} 
\begin{equation}
\hat{s}_{1} =\arg \min_{\tilde{{s}}_{1}}\text{ }\left\vert\sum_{\ell =1}^{T_{1}} {r}_{1}[\ell]- T_1 h_{1}\sqrt{\alpha _{1}}\tilde{s}%
_{1}\right\vert ^{2}.  \label{eq:MLD_SU}
\end{equation}
To detect ${s}_n$ for $n \in \{2,3,\dots,N\}$, \gls{sic} can be used recursively to eliminate the interference of higher power symbols. That is,  
\begin{equation}\label{eq:SIC general}
\acute{\mathbf{r}}_n = \mathbf{r}_n - h_n \sqrt{{\alpha _{n-1}}} \mathbf{\hat{{s}}}_{n-1}, ~n>1.
\end{equation}
However, the elements of $\mathbf{r}_n$ should be fragmented into $\delta_n$ vectors, and the data symbol in each vector is detected using a single-user \gls{mld},
\begin{equation}
\hat{s}_{n}[i]=\arg \min_{s_{n}}\left\vert \sum_{v=1+\left( i-1\right)
T_{n}}^{iT_{n}}\acute{r}_{n}[v]-T _{n}h_{n}\sqrt{\alpha _{n}}\tilde{s}%
_{n}[i]\right\vert ^{2}
\label{eq:mid user detector general}
\end{equation}
where $i=[1,2,\dots,\delta_n]$.
\section{Conditional BER Analysis ($N=3$)}\label{sec:BER Analysis}
To derive the \gls{ber} for an arbitrary number of users, we use the results in \cite{yahya2024designMR_NOMA} for the case of $N=2$, derive the analysis for $N=[3,4]$, and by induction, the \gls{ber} for an arbitrary $N$ is derived. However, the case for $N=4$ is omitted due to space constraints. Moreover, the analysis is carried out for \gls{bpsk} and then generalized to \gls{qpsk}. Furthermore, the analysis is presented for the first pattern on the right-hand side of Fig. \ref{fig:constelation421}, which consists of a total of $16$ cases, including repetitions, which are reduced to $5$ distinct cases. The analysis for the remaining cases can be derived following the same approach. 
\subsection{Conditional BER Analysis of $U_1$}
\begin{figure*}
    \centering
    \includegraphics[scale=0.4]{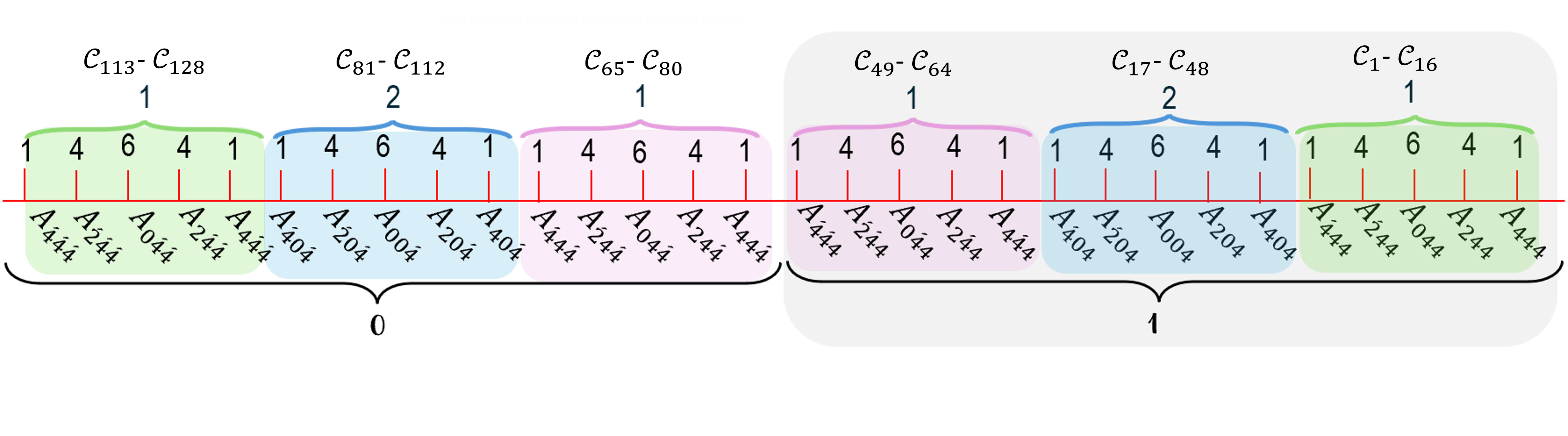}
    \vspace{-6mm}\caption{All possible values of $x^{(T_1)}_{\Sigma}$ for $U_1$ 
and their corresponding case indices $[\mathcal C_1-\mathcal C_{128}]$, $N=3$, $\boldsymbol T=$[$4$, $2$, $1$] and $\boldsymbol{M}$ = [2, 2, 2].}
    \label{fig:constelation421}
\end{figure*}Based on the \gls{mld} in \eqref{eq:MLD_SU}, the $\ell$th received sample at
the first user receiver $r_1[\ell]$ can be written as 
\begin{equation}
{r}_{1}[\ell ]=h_{1}x\left[ \ell \right] +w_{1,\ell }.
\end{equation}%
\Rthree{\label{R_3_9}Considering a three-user case,  
\begin{equation}
{x}[\ell]=\sum\nolimits_{n=1}^{3}\sqrt{\alpha _{n}}s_{n}[\ell] 
\triangleq A_{s_3[\ell]\,s_2[\ell]\,s_1[\ell]},
\label{eq:sum-U1}
\end{equation}
where $s_n[\ell]\in \{\grave{1},1\}$ and $\grave{a}\triangleq -a$. 
For sample combining at $U_1$, let 
$u_n\triangleq\sum\nolimits_{\ell=1}^{v}s_n[\ell]$, 
with $v=T_1$. Consequently,
\begin{equation}
\label{eq:x_sigma_1}
    \sum_{\ell =1}^{T_{1}}x[\ell]\triangleq x^{(T_1)}_{\Sigma}
=\sum_{\ell=1}^{T_1}A_{s_3[\ell]\,s_2[\ell]\,s_1[\ell]}
    =A_{u_3\,u_2\,u_1}
\end{equation}
where
$
A_{u_3\,u_2\,u_1}=\sum\nolimits_{n=1}^{3}\sqrt{\alpha _{n}}u_{n}
$. To illustrate the sample-combining process, Fig.~\ref{fig:constelation421}
shows the effective constellation obtained after combining the samples used
for detecting $U_1$, i.e., $x^{(T_1)}_{\Sigma}$ for all cases. For the considered
three-user example, the symbol-duration and modulation-order vectors are
$\boldsymbol{T}=[4,2,1]$ and $\boldsymbol{M}=[2,2,2]$, respectively; hence,
$T_1=4$, all users employ \gls{bpsk}, and $\boldsymbol{\delta}=[1,2,4]$. Since each symbol of $U_n$ occupies $T_n$ samples, if $k_n$ symbols out of the $\delta_n$ symbols are equal to $+1$, where
$k_n=0,1,\ldots,\delta_n$, then
$u_n=T_nk_n-T_n(\delta_n-k_n)=T_n(2k_n-\delta_n)=2T_nk_n-4$. Therefore, the possible accumulated values for
each user are $u_1\in\{\grave{4},4\}$, $u_2\in\{\grave{4},0,4\}$, and
$u_3\in\{\grave{4},\grave{2},0,2,4\}$. For example, for $U_3$,
$T_3=1$ and $\delta_3=4$, thus $k_3\in\{0,1,2,3,4\}$, which results in
$u_3\in\{\grave{4},\grave{2},0,2,4\}$. Considering all combinations of
$\{u_1,u_2,u_3\}$, $x^{(4)}_{\Sigma}$ has
\Rthree{\label{R_3_8}$\prod_{n=1}^{3}(\delta_n+1)\!\!=\!\!2\times3\times5\!\!=\!\!30$} distinct values. However, several transmitted sequences may produce the same accumulated
value of $x^{(4)}_{\Sigma}$. Specifically, in Fig.~\ref{fig:constelation421},
the main groups correspond to the possible values of $u_1$, the brackets
inside each group correspond to the possible values of $u_2$, and the ticks
inside each bracket correspond to the possible values of $u_3$. The labels
above the ticks indicate the multiplicities of the $u_3$ values, while the
numbers above the brackets indicate the multiplicities of the corresponding
$u_2$ patterns. Hence, the total repetition of each value of
$x^{(4)}_{\Sigma}$ can be $1,2,4,6,8$, or $12$.

This results in a total number of transmitted cases of
$J=2^{\sum_{i=1}^{3}\delta_i}\!\!=\!\!128$, where the $j$th case is
denoted by $\mathcal{C}_j$. In other words, the transmitted vector
$\mathbf{x}=[x[1],x[2],x[3],x[4]]^{\top}$ has $128$ possible cases.
Therefore, to account for the repeated values of $x_{\Sigma}$, an indexing
rule is needed to associate each case number $\mathcal{C}_j$ with its
corresponding transmitted sequence after repetition and, consequently, with
its signal amplitude. The mapping is obtained as follows: The values of
$x[\ell]~\forall \ell$ and $x^{(4)}_{\Sigma}$ for $\mathcal{C}_j$ are
computed by first defining an $N\times T_1$ matrix
$\mathbf{Z}\in\mathbb{R}^{N\times T_1}$, whose elements are defined by
\begin{equation}
\label{eq:Z_general}
Z_{n,\ell}
=  \sum_{m=n+1}^{N}\delta_m
  + \left\lfloor \frac{(\ell-1)\,\delta_n}{T_1} \right\rfloor
\end{equation}
where $n$ is the user index and $\ell=1,2,\dots,T_{1}$ is the sample index.
For a specific $\mathcal C_j$, $\mathbf{Z}$ can be used to find
$\mathbf s^{\mathcal{C}_j}_{n}$ $\forall n$, where the BPSK value of $U_n$
in the $\ell$th sample can be computed as  
\begin{equation}
\mathfrak{b}_{n,\ell}|\mathcal{C}_j
=2\left( \left\lfloor \frac{j-1}{2^{Z_{n,\ell}}}\right\rfloor \bmod 2 \right )-1. 
\label{eq:bit_phi}
\end{equation}
For the considered setup, $N=3$ and $\boldsymbol{T}=[4,2,1]$, thus
\begin{equation}
\mathbf{Z}=\begin{bmatrix}
6&6&6&6\\
4&4&5&5\\
0&1&2&3
\end{bmatrix}.
\label{eq:Q_rule}
\end{equation}
For example, for $\mathcal{C}_{20}$, when $\{n=1,\ell=1,2,3,4\}$, then
$Z_{1,\ell}=6~\forall \ell$, thus
$\mathfrak{b}_{1,\ell}|\mathcal{C}_{20}=\grave{1}~\forall \ell$, and
$\mathbf{s}_1^{\mathcal{C}_{20}}=[\grave{1},\grave{1},\grave{1},\grave{1}]$.
For $n=2$, $Z_{2,1}=Z_{2,2}=4$ and $Z_{2,3}=Z_{2,4}=5$. Therefore,
$\mathfrak{b}_{2,1}|\mathcal{C}_{20}=\mathfrak{b}_{2,2}|\mathcal{C}_{20}=1$
and
$\mathfrak{b}_{2,3}|\mathcal{C}_{20}=\mathfrak{b}_{2,4}|\mathcal{C}_{20}=\grave{1}$,
which gives
$\mathbf{s}_2^{\mathcal{C}_{20}}=[1,1,\grave{1},\grave{1}]$. Applying the same approach for $U_3$ gives
$\mathbf{s}_3^{\mathcal{C}_{20}}=[1,1,\grave{1},\grave{1}]$.
Thus, $u^{\mathcal{C}_{20}}_1=\grave{4}$,
$u^{\mathcal{C}_{20}}_2=0$, and $u^{\mathcal{C}_{20}}_3=0$. Finally,
$x^{(4)}_{\Sigma}|\mathcal{C}_{20}
=A_{u^{\mathcal{C}_{20}}_3\,u^{\mathcal{C}_{20}}_2\,u^{\mathcal{C}_{20}}_1}
=A_{00\grave{4}}$}.


Note that $s_{n}$ is \gls{iid}, hence $\mathcal{C}_j$ are equiprobable $\forall j$, and the \gls{ber} of $U_1$ for a given $h_1$ is computed by considering $\mathcal{C}_j~\forall j$,
\begin{equation} \label{generalprobability of error}
    P_{B_1} =\sum_j \Pr(E_1|\mathcal{C}_j) \Pr(\mathcal{C}_j)=\frac{1}{128} \sum\nolimits_{j=1}^{128}  \Pr(E_1|\mathcal{C}_j).
\end{equation}
where $\Pr(E_1|\mathcal{C}_j)=P_{B_1}|\mathcal{C}_j$. The list of all possible cases is given as follows:

$\mathcal{C}_1$: $\mathbf{x} = \left[ A_{111}, A_{111}, A_{111}, A_{111} \right]^\top$ and $ x^{(4)}_{\Sigma}=A_{444}$. Subsequently, after phase equalization, the combined received signal $\sum_{\ell=1}^{T_1}r_1[\ell]e^{-j\theta_1}=\dot{r}_1= \mathcal{A}^1_{444}+\sum_{i=1}^4 w_{1,i}$, 
where $h_1=|h_1|e^{-j\theta}$. For circularly symmetric \gls{awgn} and \gls{iid} \(w_{1,v}\), $0<v\leq T_1$, 
$w_1^{(V)} \triangleq \sum_{v=1}^{V} w_{1,v} \sim \mathcal{CN}\!\left(0,\, V\sigma_w^{2}\right)$, where $V\in\{T_1, T_2,\dots,T_N\}$ is the period of the symbol to be detected. Accordingly, and defining  $|h_i| A_{abc}\triangleq \mathcal{A}^i_{abc}$
\begin{align}
    \Pr(E_1|\mathcal{C}_1)=& \Pr( \mathcal{A}^1_{444} + w_1^{(4)} \leq 0)\notag\\
    =& \Pr(w_1^{(4)} \geq  \mathcal{A}^1_{444}) \notag\\
    =& \frac{1}{2\sqrt{2\pi\sigma_w^2}} \int_{ \mathcal{A}^1_{444}}^{\infty} e^{-\frac{t^2}{8\sigma_w^2}} dt
\end{align}
which can be evaluated to,
\begin{equation}
     \Pr(E_1|\mathcal{C}_1) =Q\left(A_{444}\sqrt{\frac{\gamma_1}{4}}\right)=Q\left(A_{444}G_{1,1}\right)
     \label{Pb1Case1}
\end{equation}
where $G_{m,n}=\sqrt{\frac{\gamma_m}{T_n}}$, $\gamma_m = \tfrac{|h_m|^2}{\sigma_{\mathrm{w}}^2}$, and $Q(\cdot)$ denotes the complementary \gls{cdf} of the standard Gaussian random variable. Following the same procedure used with $\mathcal{C}_1$, the expressions for the remaining cases are:

$\mathcal{C}_2$: $\mathbf{x} = \left[ A_{\grave111}, A_{111}, A_{111}, A_{111} \right]^\top$, $x^{(4)}_{\Sigma}= A_{244}$, and
\begin{equation}
\Pr(E_1|\mathcal{C}_2) = \Pr \left( w_1^{(4)} < -\mathcal{A}^1_{244} \right) = Q \left( A_{244} G_{1,1} \right).
\label{Pb1Case2}
\end{equation}
For $\{\mathcal{C}_3, \mathcal{C}_4,\mathcal{C}_5\}$, the respective values of $\mathbf{x}$ are $ \left[ A_{111}, A_{\grave111}, A_{111}, A_{111} \right]^\top$, $\left[ A_{111}, A_{111}, A_{\grave111}, A_{111} \right]^\top$, and $\left[ A_{111}, A_{111}, A_{111}, A_{\grave111} \right]^\top$. Thus $x^{(4)}_{\Sigma}$ for these cases is equal to $\mathcal{C}_2$.

$\mathcal{C}_6$: $\mathbf{x} = \left[ A_{\grave111}, A_{\grave111}, A_{111}, A_{111} \right]^\top$, $x^{(4)}_{\Sigma} = A_{044}$, and
\begin{equation}
\Pr(E_1|\mathcal{C}_6) = \Pr \left( w_1^{(4)} < -\mathcal{A}^1_{044} \right) = Q \left( A_{044} G_{1,1} \right).
\label{Pb1Case6}
\end{equation}

$\mathcal{C}_7,\mathcal{C}_8,\ldots ,\mathcal{C}_{11}$, $\mathbf{x} = \left[ A_{111}, A_{\grave111}, A_{111}, A_{\grave111} \right]^\top$, $\left[ A_{111}, A_{\grave111}, A_{\grave111}, A_{111} \right]^\top$, $\left[ A_{\grave111}, A_{111}, A_{111}, A_{\grave111} \right]^\top$, $\left[ A_{\grave111}, A_{111}, A_{\grave111}, A_{111} \right]^\top$, and $\left[ A_{111}, A_{111}, A_{\grave111}, A_{\grave111} \right]^\top$, respectively. Therefore, $x^{(4)}_{\Sigma}$ is equal to $\mathcal{C}_6$.

$\mathcal{C}_{12}$: $\mathbf{x} = \left[ A_{\grave111}, A_{\grave111}, A_{\grave111}, A_{111} \right]^\top$, $x^{(4)}_{\Sigma} = A_{\grave244}$, and,
\begin{align}
\Pr(E_1|\mathcal{C}_{12}) = \Pr \left( w_1^{(4)} < -\mathcal{A}^1_{\grave244} \right)= Q \left( A_{\grave244} G_{1,1} \right).
\end{align}

$\mathcal{C}_{13}$, $\mathcal{C}_{14}$, $\mathcal{C}_{15}$: $\mathbf{x} = \left[ A_{111}, A_{\grave111}, A_{\grave111}, A_{\grave111} \right]^\top$, $\left[ A_{\grave111}, A_{111}, A_{\grave111}, A_{\grave111} \right]^\top$, and $\left[ A_{\grave111}, A_{\grave111}, A_{111}, A_{\grave111} \right]^\top$, respectively, $x_{\Sigma}$ for all cases are all similar to $\mathcal{C}_{12}$.

$\mathcal{C}_{16}$: $\mathbf{x} = \left[ A_{\grave111}, A_{\grave111}, A_{\grave111}, A_{\grave111} \right]^\top$, $x^{(4)}_{\Sigma} = A_{\grave444}$, and,
\begin{equation}
\Pr(E_1|\mathcal{C}_{16}) = \Pr \left( w_1^{(4)} < -\mathcal{A}^1_{\grave444} \right) = Q \left( A_{\grave444} G_{1,1} \right).
\end{equation}
\Rthree{$[R_{3,10}]$\label{R_3_10}
As shown in Fig.~\ref{fig:constelation421}, BPSK sign symmetry pairs
$\mathcal{C}_{q}$ with $\mathcal{C}_{129-q}$, $q=1,\ldots,64$, with
identical conditional error probabilities. Moreover, the corresponding
cases in $\mathcal{C}_{17}$--$\mathcal{C}_{32}$ and
$\mathcal{C}_{33}$--$\mathcal{C}_{48}$ are equiprobable. Therefore,
$\mathcal{C}_{1}$--$\mathcal{C}_{64}$ reduce to three probability groups,
each containing five distinct terms, yielding $3\times5=15$ terms.
Accounting for their multiplicities and symmetric counterparts, the
conditional \gls{ber} of $U_1$ for the symbol durations $[4,2,1]$ is}
\begin{equation}
\!\!\!P_{B_{1}}\!=\!\frac{%
1}{64}\sum_{j=1}^{3}\tbinom{2}{j-1}\sum_{i=1}^{5}\tbinom{4}{i-1}Q\left( A_{E^{(T_3,T_1)}_{i}E^{(T_2,T_1)}_{j}4}G_{1,1}\right)
\label{eq:FarUserBER421}
\end{equation}
 where, $[-T_1,-T_1+2T_n,\dots,T_1]\triangleq \boldsymbol{E}^{(T_n,T_1)}$. For $\boldsymbol T=[4,2,1]$, $\boldsymbol{E}^{(T_2,T_1)}=\boldsymbol{E}^{(2,4)}=[\grave{4},0,4]$, thus, $E^{(2,4)}_1=\grave{4}$, $E^{(2,4)}_2=0$, and $E^{(2,4)}_3=4$.
 
For \gls{qpsk}, which can be viewed as two orthogonal \gls{bpsk} symbols, the  \gls{ber} equations preserve the general structure of \eqref{eq:FarUserBER421} \cite{assaf2020noma}, with the modification that $\bar{A}_{(.)}$ is used instead of $A_{(.)}$, where $\bar{A}_{s_3\,s_2\,s_1}=\frac{A_{s_3\,s_2\,s_1}}{\sqrt{\kappa_n}}$, and ${\kappa_n} = 2\frac{(M_i-1)}{3} \forall \, n\in N$. It is worth mentioning that this applies to all forthcoming derived \gls{ber} expressions presented in this work when \gls{qpsk} modulation is considered. 
\subsection{Conditional BER Analysis of $U_2$}
The detection of $U_2$ is performed by detecting $U_1$ symbol first, applying \gls{sic} \eqref{eq:SIC general}, and finally applying the \gls{mld} \eqref{eq:mid user detector general}. However, because $U_2$ has two symbols in the transmission window, the \gls{mld} is applied once for each symbol. Therefore, the two symbols of $U_2$ are detected independently and have equal error probability. Hence, it is sufficient to derive the \gls{ber} only for one of them, and the time index for $s_1$ and $s_2$ can be dropped. The \gls{ber} of $U_2$, conditioned on the detection outcome of $U_1$, is given by, \begin{equation}
P_{B_{2}}=\sum_{j}\sum_{\nabla^1 }\Pr \left( \mathrm{Er}_2,\nabla^1 ,%
\mathcal{C}_{j}\right) 
\end{equation}%
where $\mathrm{Er}_2$ represents the event $\hat{s}_{2}\neq s_{2}$, $\nabla^1 \in \left\{ \nabla^1_{S},\nabla^1 _{F}\right\} $ is the \gls{sic} 
process indicator, $\nabla^1_{S}\triangleq \hat{s}_{1}=s_{1}$ and $\nabla^1
_{F}\triangleq \hat{s}_{1}\neq s_{1}$. Because $\Pr \left( \mathcal{C}%
_{j}\right) $ are equiprobable $\forall j$, then $P_{B_{2}}$ can be written
as 
\begin{align}
P_{B_{2}}& =\sum_{j}\sum_{\nabla ^{1}}\Pr \left( \mathrm{Er}_{2}
|[\nabla ^{1},\mathcal{C}_{j}]\right) \Pr \left( \nabla ^{1},\mathcal{C}%
_{j}\right)   \notag \\
& =\sum_{j}\sum_{\nabla ^{1}}\Pr \left( \mathrm{Er}_{2}|[\nabla ^{1},\mathcal{%
C}_{j}]\right) \Pr \left( \nabla ^{1}|\mathcal{C}_{j}\right) \Pr \left( 
\mathcal{C}_{j}\right)   \label{eq:general:U2 BER}
\end{align}
where in this example $\Pr\left(\mathcal{C}_{j}\right)=1/128$. 

\color{black}

Based on \eqref{eq:general:U2 BER}, the per-case error for the configuration $\boldsymbol T=[4,2,1]$ for the same cases applied for $U_1$, $\mathcal C_1$-$\mathcal C_{5}$ is obtained as follows,\newline
   $\mathcal{C}_1$: Under this condition, the bit error probability for detecting $U_1$ symbols at $U_2$ follows the same expression as in~\eqref{Pb1Case1}, except that $G_{1,1}$ is replaced by $G_{2,1}$ to
account for $\gamma_2$. Therefore, $\Pr(\nabla^1_F|\mathcal{C}_1) = Q\left(A_{444} G_{2,1}\right)
$, which is also applied in all
relevant cases. In the case of $\nabla_F^1$, the \gls{sic} is performed as, $[A_{111},A_{111},A_{111},A_{111}]^\top-[A_{001},A_{001},A_{001},A_{001}]^\top=[A_{110},A_{110},A_{110},A_{110}]^\top$. Subsequently, to detect $\{s_{2}[1],s_{2}[2]\}$, the values of $x^{(T_2)}_{\Sigma}$ over the two symbols are $[A_{220},A_{220}]$. Furthermore, since $\{s_{2}[1],s_{2}[2]\}$ are
independent and detected independently, they experience similar error probabilities. Consequently, it is sufficient to consider only the $s_{2}[1]$. Hence,
\begin{equation}
\!\!\!\Pr(\mathrm{Er}_2|[\nabla^1_S,\mathcal{C}_1])\!=\!\Pr \big(\mathcal{A}^2_{220}\!+\!w^{(T_2)}_{2}\!\! <\!0\big)\!=\! Q\left(A_{220} G_{2,2}\right).
\label{eq:Pr_E2_S_C1}
\end{equation}
Following the same procedure for $\nabla_F^1$,  $[A_{111},A_{111},A_{111},A_{111}]^{\top}-[A_{00\grave1},A_{00\grave1},A_{00\grave1},A_{00\grave1}]^{\top}\!\!=\!\![A_{112},A_{112},A_{112},A_{112}]^{\top}$, thus, $x^{(2)}_{\Sigma}$ for the two symbols are $[A_{224},A_{224}]^{\top}$. Consequently,
\begin{equation}
\!\!\!\Pr(\mathrm{Er}_2|[\nabla^1_F,\mathcal{C}_1])\!\! =\!\! \Pr \big(\!\mathcal{A}^2_{224}+w^{(T_2)}_{2}\!<\! 0\big)\!=\!\! Q\left(A_{224} G_{2,2}\right).
\end{equation}

Following the same procedure, the error probabilities for $\mathcal{C}_2$, $\mathcal{C}_3$, $\mathcal{C}_4$ and $\mathcal{C}_5$ are:
\begin{equation}
\!\!\Pr(\mathrm{Er}_2|[\nabla^1_S,\mathcal{C}_2])\!=\!\Pr \big( \mathcal{A}^2_{020}+w^{(2)}_{2}\!<\! 0\big)\!=\!Q\left(A_{020} G_{2,2}\right)
\label{eq:PB2C2_1}
\end{equation}
\begin{equation}
\Pr(\mathrm{Er}_2|[\nabla^1_F,\mathcal{C}_2])\!\! =\!\!  \Pr\big(\mathcal{A}^2_{024}+w^{(2)}_{2}\!\!<0 \big)\!\!=\!\!Q\left(A_{024} G_{2,2}\right).
\label{eq:PB2C2_2}
\end{equation}
\begin{equation}
\Pr(\mathrm{Er}_2|[\nabla^1,\mathcal{C}_3])\!\!=\Pr(\mathrm{Er}_2|[\nabla^1,\mathcal{C}_2]), \nabla^1\in \{ \nabla^1_S, \nabla^1_F\}.
\end{equation}
    \begin{equation}
\Pr(\mathrm{Er}_2|[\nabla^1_S,\mathcal{C}_4])\!=\!\Pr\! \big(\!\mathcal A^2_{220}\!+w^{(2)}_{2}\!\!<0\!\big)\!=\!Q\left(A_{220}G_{2,2}\right).
\label{eq:PB2C4_1}
\end{equation}
\begin{equation}
\!\!\Pr(\mathrm{Er}_2|[\nabla^1_F,\mathcal{C}_4])\!=\!\Pr \big(\mathcal A^2_{224}\!+\!w^{(2)}_{2}\!<\!\! 0\big)\!=\!\!Q\left(\!A_{224} G_{2,2}\!\right)\!.
\label{eq:PB2C4_2}
\end{equation}
 Finally, $\Pr(\mathrm{Er}_2|[\nabla^1_S,\mathcal{C}_5])$ and $\Pr(\mathrm{Er}_2|[\nabla^1_F,\mathcal{C}_5])$ are similar to those in \eqref{eq:PB2C4_1} and \eqref{eq:PB2C4_2}, respectively.

By following the same approach for the remaining cases,  
\begin{multline}
\label{eq:U2 BER 4:2:1 (term1)}
\Pr \left(
\mathrm{Er}_2,\nabla^1_F \right)
= \frac{1}{64}
\Bigg[
\sum_{k=1}^{2}\sum_{j=1}^{3}\sum_{i=j}^{j+2} \binom{2}{j-1}
\\ \times 
  \Big(
Q(A_{I_j^{2}\,2\,4})\,Q(A_{E_i^{(1,4)}\,E_{k+1}^{(2,4)}\,4})
-
Q(A_{I_j^{2}\,\grave{2}\,4})\, \\ \times Q(A_{E_i^{(1,4)}\,E_{k}^{(2,4)}\,4})
\Big)
+
\sum_{l\in\{\grave{4},\,0\}}\sum_{i=1}^{5}\binom{4}{i-1} Q(A_{E_i^{(1,4)}\,l\,4})
\Bigg].
\end{multline}
By induction, the expression of $\Pr \left(
\left[\mathrm{Er}_2,\nabla^1_F \right] \right)$ is generalized for arbitrary $\boldsymbol{T}$ in \eqref{eq:U_2 BER exact(term1) any symbol duration}, where $\boldsymbol{I}^{T_n}=[-T_n,-T_n+2,\dots, T_n]$, for example for $T_2=2$ then $\boldsymbol{I}^{T_2}=\boldsymbol{I}^{2}=[\grave{2}, 0, 2]$, where $\grave2\triangleq-2$. Also, $I^{2}_1=\grave{2}$, $I^{2}_2=0$, and $I^{2}_3=2$.
\begin{figure*}
\begin{multline}\label{eq:U_2 BER exact(term1) any symbol duration}
\Pr \left( %
 \mathrm{Er}_2,\nabla^1 _F \right) =\frac{1}{2^{\delta
_{2}+\delta_3}}\Bigg[\sum_{k=1}^{\delta
_{2}}\sum_{j=1}^{T_{1}-T_{2}+1}\sum_{i=j}^{T_{1}-T_{2}+j}\tbinom{\delta
_{2}-1}{k-1}\tbinom{T_{1}-T_{2}}{j-1} 
 \bigg[ Q\left(
A_{I_{j}^{T_{2}}\, T_{2}\,2T_{2}}G_{2,2}\right)Q\left( A_{E_{i}^{(T_{3},T_1)}E_{k+1}^{(T_{2},T_1)}T_{1}}G_{2,1}\right)
\\-Q\left(
A_{I_{j}^{T_{2}}\,-T_{2}\,2T_{2}}G_{2,2}\right)Q\left( A_{E_{i}^{(T_{3},T_1)}E_{k}^{(T_{2},T_1)}\,T_{1}}G_{2,1}\right) \bigg]
+\sum_{l=1}^{\delta _{2}}\sum_{i=1}^{\delta _{3}+1}\tbinom{\delta _{2}-1}{l-1%
}Q\left( A_{E_{i}^{(T_{3},T_1)}\,E_{l}^{(T_{2},T_1)}\,T_{1}}G_{2,1}\right) \Bigg].
\end{multline}
---------------------------------------------------------------------------------------------------------------------------------------------------------
\end{figure*}
The term that corresponds to $\nabla_S^1$ is given by,
\begin{multline}
\Pr \left( \mathrm{Er}_{2},\nabla _{S}^{1}\right) =\sum_{i=1}^{3}\tbinom{2}{%
i-1}Q(A_{I_{i}^{2}\,2\,0})  \label{eq:U2 BER 4:2:1  (term2)} \\
\times \bigg[1-\frac{1}{64}\bigg(\sum_{k=1}^{2}\sum_{j=4-i}^{6-i}\tbinom{1}{%
k-1}\tbinom{2}{j-2-i}Q(A_{E_{j}^{(1,4)}\,E_{k}^{(2,4)}\,4}) \\
+\sum_{k=2}^{3}\sum_{j=i}^{i+2}\tbinom{1}{k-2}\tbinom{2}{j-i}%
Q(A_{E_{j}^{(1,4)}\,E_{k}^{(2,4)}\,4})\bigg)\bigg].
\end{multline}
For arbitrary $\boldsymbol{T}$, the  generalization of \eqref{eq:U2 BER 4:2:1  (term2)} is given in \eqref{eq:U2 BER (term2) any symbol duration}.
Hence, $P_{B_2}$ for $N=3$ and $\boldsymbol{T}=[4, 2, 1]$ is the sum of \eqref{eq:U2 BER 4:2:1 (term1)} and \eqref{eq:U2 BER 4:2:1  (term2)}, while for arbitrary $\boldsymbol{T}$ it is the sum of \eqref{eq:U_2 BER exact(term1) any symbol duration} and \eqref{eq:U2 BER (term2) any symbol duration}.
\begin{figure*}
\begin{multline}\label{eq:U2 BER (term2) any symbol duration}
\Pr \left( \mathcal{C}_{j}\right) \sum_{j}\Pr \left( %
\left[ \mathrm{Er}_2,\nabla^1_S \right] |\mathcal{C}_{j}\right) =\sum_{i=1}^{T_{2}+1}\tbinom{T_{2}}{i-1}Q\left(
A_{I_{i}^{T_{2}}T_{2}0}G_{2,2}\right)\Bigg[1-\frac{1}{2^{\delta_2+\delta_3}} \Bigg(\sum_{k=1}^{\delta
_{2}}\sum_{j=T_{2}+2-i}^{T_{1}+2-i}\tbinom{\delta _{2}-1}{k-1}\tbinom{%
T_{1}-T_{2}}{j-T_{1}+T_{2}-i}\\ \times Q\left( A_{E_{j}^{(T_{3},T_{1})}E_{k}^{(T_{2},T_1)}T_{1}}%
G_{2,1}\right)  
+\sum_{k=2}^{\delta _{2}+1}\sum_{j=i}^{T_{1}-T_{2}+i}\tbinom{\delta _{2}-1}{%
k-2}\tbinom{T_{1}-T_{2}}{j-i}Q\left( A_{E_{j}^{(T_{3},T_1)}E_{k}^{(T_{2},T_1)}T_{1}}%
G_{2,1}\right) \Bigg)\Bigg].
\end{multline}
-----------------------------------------------------------------------------------------------------------------------------------------------------------
\end{figure*}
\subsection{Conditional BER Analysis of $U_{3}$}
Based on the receiver design in \eqref{eq:mid user detector general}, $U_{3}$ performs two \gls{sic} operations, and hence its \gls{ber} depends on the outcome of both \gls{sic} operations. Because all symbols of $U_3$ have identical error probability and are detected independently, it is sufficient to consider only the first symbol of $U_3$. Therefore, for $N=3$,
\begin{equation}
    P_{B_{3}}= \sum_{j}\sum_{\nabla^1 }\sum_{\nabla^2 }\Pr \left( %
 \mathrm{Er}_3, \nabla^2,\nabla^1,  \mathcal{C}_{j}\right). 
\end{equation}
Using the chain rule, $P_{B_3}$ can be written as,
\begin{multline}
P_{B_{3}} 
 = \sum_{j}\sum_{\nabla^1 }\sum_{\nabla^2 }\Pr \left( %
 \mathrm{Er}_3|[\nabla^2,\nabla^1,  \mathcal{C}_{j}]\right) \Pr \left( \nabla^2|[\nabla^1,  \mathcal{C}_{j}]\right)
  \\ \times \Pr \left( \nabla^1|\mathcal{C}_{j}\right)\Pr \left( \mathcal{C}_{j}\right)
\label{eq:general:U3 BER}
\end{multline}
where $\mathrm{Er}_3$ represents the event $\hat{s}_{3}[1]\neq s_{3}[1]$, $\nabla^2 \in \left\{ \nabla^2_{S},\nabla^2_{F}\right\} $, $\nabla^2_{S}\triangleq \hat{s}_{2}[1]=s_{2}[1]$, and $\nabla^2
_{F}\triangleq \hat{s}_{2}[1]\neq s_{2}[1]$. Similar to the derivation of $P_{B_1}$ and $P_{B_2}$, $P_{B_3}$ is derived as the average of all possible data transmission cases.

In \eqref{eq:general:U3 BER}, $\Pr(\mathcal{C}_j)=1/128$ and $\Pr(\nabla^1_F \!\mid \mathcal{C}_j)$ are equivalent to those derived for $U_2$ with the exception that $G_{2,1}$ is replaced with $G_{3,1}$. 
For example, 
$\Pr(\nabla^1_F \!\mid \mathcal{C}_1) = Q\!\left(A_{444} G_{3,1}\right)
$. The term $\Pr \left( \nabla^2|[\nabla^1, \mathcal{C}_{j}]\right)$ can be evaluated by noting that 
$\Pr\left(\nabla_F^2|[\nabla^1, \mathcal{C}_{j}]\right)=\Pr(\mathrm{Er}_2| [\nabla^1, \mathcal{C}_j])$, which are given in \eqref{eq:Pr_E2_S_C1} and the subsequent formulae, except that $G_{2,2}$ is replaced  by $G_{3,2}$.
For example, $\Pr(\nabla_F^2|[\nabla^1_S,\mathcal{C}_1]) =Q\left(A_{220} G_{3,2}\right)$, and $\Pr(\nabla_F^2|[\nabla^1_F,\mathcal{C}_1]) =Q\left(A_{224} G_{3,2}\right)$. 
By considering the four possible scenarios of $\nabla^1$ and $\nabla^2$ we obtain, 
\begin{align}
\Pr(\mathrm{Er}_3|[\nabla^2_S,\nabla^1_S,\mathcal{C}_1]) 
  &=\Pr\left(\mathcal A^3_{{1}00}+w_{3,1}<0\right)\notag \\
&= Q\!\left(A_{100} G_{3,3}\right).
\end{align}
\begin{align}
\Pr(\mathrm{Er}_3|[\nabla^2_F,\nabla^1_S,\mathcal{C}_1]) 
  &=\Pr\left(\mathcal A^3_{120}+w_{3,1}<0\right)\notag \\
&= Q\!\left(A_{120} G_{3,3}\right).
\end{align}
\begin{align}
\Pr(\mathrm{Er}_3|[\nabla^2_S,\nabla^1_F,\mathcal{C}_1]) 
  &=\Pr\left(\mathcal A^3_{102}+w_{3,1}<0\right) \notag \\
&= Q\!\left(A_{102} G_{3,3}\right).
\end{align}
\begin{align}
\Pr(\mathrm{Er}_3|[\nabla^2_F,\nabla^1_F,\mathcal{C}_1]) 
  &=\Pr\left(\mathcal A^3_{122}+w_{3,1}<0\right)\notag \\
&= Q\!\left(A_{122} G_{3,3}\right).
\end{align}
Finally, the resulting expression for $\mathcal C_1$ is given by,
\begin{multline}
\sum_{\nabla^1 }\sum_{\nabla^2 }\Pr \left( %
[\mathrm{Er}_3,\nabla^2,\nabla^1] |\mathcal C_1\right)=F\left(A_{444}G_{3,1}\right)F\left(A_{220} G_{3,2}\right)\\ \times  Q\left(A_{100}G_{3,3}\right)+F\left(A_{444}G_{3,1}\right) Q\left(A_{220} G_{3,2}\right)Q\left(A_{120}G_{3,3}\right) \\+ Q\left(A_{444}G_{3,1}\right) F\left(A_{224}G_{3,2}\right)  Q\left(A_{102}G_{3,3}\right) + Q\left(A_{444}G_{3,1}\right)\\ \times Q\left(A_{224}G_{3,2}\right) Q\left(A_{122}G_{3,3}\right).
\end{multline}

Following the same approach for the remaining cases and after some manipulation, \eqref{eq:general:U3 BER} is evaluated in Appendix \ref{app:U3-BER} as the sum of \eqref{eq:user1correct,user2correct_symbols(a,z,q)}, \eqref{eq:user1correct,user2wrong_symbols(a,z,g)}, \eqref{eq:user1wrong,user2correct_symbol}, and \eqref{eq:user1wrong,user2wrong_symbols(a,b,c)}. 
\section{Conditional \Gls{ber} Analysis $(N>3)$}\label{sec:BER Analysis_N>3}
\subsection{BER Analysis of $U_1$}
\Rthree{\label{R_3_11} By induction on $N=2$ provided in \cite{yahya2024designMR_NOMA}, and $N=3$ cases, it can be observed that there is a consistent combinatorial regularity in the superposed constellation.} Specifically, as either the number of users, or the symbol‑duration ratios increase, the multiplicity of identical signal points grows according to Pascal’s triangle, i.e., the binomial coefficients. Recognizing such a structure obviates case‑by‑case derivations and enables the generalization of $U_1$ \gls{ber} for $N>3$ as  
\begin{multline}
\label{eq:U_1 BER for arbitrary symbol duration and number of users}
    P_{B_{1}}=\frac{1}{2^{\sum^N_{j=1}\delta_j}}\Bigg(\sum_{k=1}^{\delta_1+1}\tbinom{\delta_1}{k-1}\sum_{l=1}^{\delta_2+1}\binom{\delta_2}{l-1}\sum_{i=1}^{\delta_3+1}\tbinom{\delta_3}{i-1}\\ \dots\sum_{j=1}^{\delta_N+1}\tbinom{\delta_N}{j-1}Q\left(A_{E^{(T_N,T_1)}_j\dots\,  E^{(T_3,T_1)}_iE^{(T_2,T_1)}_lE^{(T_1,T_1)}_k}G_{1,1}\right)\Bigg)
\end{multline}
\RTwo{\label{R_2_9} As can be noted from \eqref{eq:U_1 BER for arbitrary symbol duration and number of users}, the number of terms to be computed grows exponentially versus $N$. Nevertheless, the values of $N$ are typically small to enable efficient user pairing and power optimization, and to keep the interference manageable. It is also worth noting that for $N \geq 4$, $\alpha_N \ll \alpha_{N-3}$, which implies that accurate \gls{ber} approximations can be obtained by considering $\{\alpha_N, \alpha_{N-1},\dots, \alpha_{N-2}\}=0$.}

\subsection{\gls{ber} Analysis of $U_m$ for  $m>1$}
Based on the derivations for $N$$=$$3$, the \gls{ber} expressions derived using imperfect \gls{sic} involve a large number of terms when $m$$>$$2$, due to the large number of possible cases. Therefore, considering all cases for $N>3$ analytically is intractable. To keep the analysis tractable, we adopt the perfect \gls{sic} assumption for $U_m$ with $N>3$. It is worth noting that, unlike conventional \gls{sr}-\gls{noma}, the \gls{ber} of \gls{mr}-\gls{noma} with the perfect \gls{sic} assumption closely approaches the exact \gls{ber}. Using 
\eqref{eq:FarUserBER421}, the generalized \gls{ber} expression of $U_m$ for an arbitrary number of users with arbitrary symbol durations,is given in \eqref{eq:BER_any_user}.
\begin{figure*}
\begin{equation}
\label{eq:BER_any_user}
\!\!\!P_{B_{m}}=\frac{1}{2^{%
\sum_{j=m}^{N}\delta_{j}}}\sum_{k=1}^{\frac{T_{m}}{T_{m}}+1}\sum_{l=1}^{\frac{%
T_{m}}{T_{m+1}}+1}\dots \sum_{j=1}^{\frac{T_{m}}{T_{N}}+1}\tbinom{\frac{T_{m}%
}{T_{m}}}{k-1} \tbinom{\frac{T_{m}}{T_{m+1}}}{l-1} \tbinom{\frac{T_{m}}{T_{N}}%
}{j-1}Q\left( A_{E_{j}^{(T_{N},T_m)}\dots
E_{i}^{(T_{m+2},T_m)}E_{l}^{(T_{m+1},T_m)}E_{k}^{(T_{m},T_m)}}G_{m,m}\right). \end{equation} 
\hrule
\end{figure*}
It is worth noting that the analysis of the considered perfect \gls{sic} assumes that the noise is Gaussian and the data symbol distribution is uniform \cite{al2023ber}.

\Rone{\label{R_1_4}The developed \gls{ber} analysis is also applicable to several multi-antenna configurations. For example, consider a V-BLAST \gls{mimo} with $K$ transmit/receive antennas serving $K$ \gls{mr}-\gls{noma} streams $x_k[\ell], k\in\{1,2,\dots,K\}, \ell\in\{1,2,\dots,\delta_N\}$. With \gls{zf} equalization, assuming that the channel is fixed over $T_1$, the received signal at $U_n$ can be
written as
\begin{equation}
\mathbf{r}_n[\ell]=
\mathbf{H}_n\mathbf{x[\ell]}
+
\mathbf{w}_n[\ell],
\end{equation}
where $\mathbf{r}_n\in\mathbb{C}^{K\times 1}$,
$\mathbf{H}_n\in\mathbb{C}^{K\times K}$ is the \gls{mimo} channel matrix
at $U_n$, $\mathbf{x}\in\mathbb{C}^{Kt\times 1[\ell]}$ is the $K$ transmitted
V-BLAST symbols vector, and $\mathbf{w}_n[\ell]\in\mathbb{C}^{K\times 1}$
is the \gls{awgn} vector. Applying a \gls{zf} equalizer yields,
\begin{equation}
\tilde{\mathbf{x}}[\ell]
=(\mathbf{H}_n^{H}\mathbf{H}_n)^{-1}
\mathbf{H}_n^{H}\mathbf{r}_n[\ell]
=\mathbf{x}[\ell]+\tilde{\mathbf{w}}_n[\ell],
\end{equation}
where
$\tilde{\mathbf{w}}_n=
(\mathbf{H}_n^{H}\mathbf{H}_n)^{-1}
\mathbf{H}_n^{H}\mathbf{w}_n
$ is the post-\gls{zf} noise vector and $(\cdot)^H$ denotes the Hermitian transpose operation. Separating the $k$th stream and augmenting the $\delta_N$ samples gives a formula similar to \eqref{eq:Received Signal model}, except that the noise is $\tilde{\mathbf{w}}_n$, which corresponds to a non-Gaussian noise \cite{Tessema}. Interestingly, the \gls{pdf} of the noise term $\tilde{\mathbf{w}}_n$ is equivalent to the \gls{pdf} that results from dividing $w_n$ by $h_n$ in \eqref{eq:Received Signal model}, which implies that the probability analysis for the \gls{siso} applies to the considered \gls{mimo} configuration.}

\section{Average BER Analysis in Nakagami Fading}\label{Sec:Average BER}
The average \gls{ber} for \gls{mr}-\gls{noma} operating over Nakagami-$\frak{m}$ fading channel is derived for $N=2$ in \cite{SumRateMax_MR_NOMA}. Therefore, this work extends the analysis to consider the general case of $N$ users. Specifically, we develop a closed-form expression for the average \gls{ber} of \gls{mr}-\gls{noma} over a Nakagami-$\frak{m}$ channel with an arbitrary number of users. The Nakagami-$\frak{m}$ fading model is adopted for its flexibility in representing various fading conditions. In Nakagami-$\frak{m}$ fading, the signal envelope $|h|^2\triangleq \lambda$ is defined as,
\begin{equation}
\begin{aligned}
    f(\lambda) = \frac{2\frak{m}^{\frak{m}} \lambda^{2\frak{m}-1}}{\Omega^{\frak{m}} \Gamma(\frak{m})} \ \mathrm{e}^{-\frac{\frak{m} \lambda^2}{\Omega}}
    \end{aligned}
    \label{NakagamiPDF}
\end{equation}
 where $\Omega=\mathbb{E}[\lambda^2]$, $\mathbb{E}[\cdot]$ is the expected value, and $\Gamma(\cdot)$ is the Gamma function. Therefore, the average $P_{B_m}$ is expressed as,
 \begin{equation}
     \bar{P}_{B_{m}}=\int_0^{\infty} P_{B_{m}}~ f(\lambda) d\lambda.
     \label{eq: ber_av}
 \end{equation}
For the case of integer $\frak{m}$, and by noting that \eqref{eq:U_1 BER for arbitrary symbol duration and number of users} is a single $Q$-function, \eqref{eq: ber_av} is evaluated as follows \cite{SumRateMax_MR_NOMA},
\begin{equation}
\label{eq: singleQ}
\chi\left(B, \frak{m}\right)= \int_0^{\infty} Q\left(\lambda\sqrt{B}\right) f(\lambda) d\lambda \\
\end{equation}
which, after evaluating the integral, is given by, 
\begin{equation}
\chi \left( B,\frak{m}\right) =\frac{1}{8}-\frac{1}{8}\sum_{k=0}^{\frak{m}-1}\frac{%
(2k)!}{(k!)^{2}4^{k}}\frac{\sqrt{B/2\frak{m}}}{\left( 1+B/2\frak{m}\right) ^{k+\frac{1}{2%
}}}
\label{eq: solvingsingleQ}
\end{equation}
\begin{multline} 
\!\!\! \bar{P}_{B_{m}}\!=\! \frac{1}{2^{\sum^N_{j=m}\delta_j}}\Bigg[\sum_{k=1}^{\frac{T_m}{T_m}+1}\tbinom{\frac{T_m}{T_m}}{k-1}
\sum_{l=1}^{\frac{T_m}{T_{m+1}}+1}\tbinom{\frac{T_m}{T_{m+1}}}{l-1}\dots\! \! \sum_{j=1}^{\frac{T_m}{T_N}+1}\tbinom{\frac{T_m}{T_N}}{j-1}  \\
 \times \chi \left(A^2_{E^{(T_N,T_m)}_j\,\dots \, \,E^{(T_{m+1},T_m)}_l\,E^{(T_m,T_m)}_k} \frac{d_{m}^{-\eta}}{\delta_m\sigma_w^2},\frak{m}\right)\Bigg]. \label{eq:U_1 Average BER for arbitrary symbol duration and number of users} 
\end{multline} 

\begin{figure*}[tb]
    \centering         \includegraphics[scale=0.65]{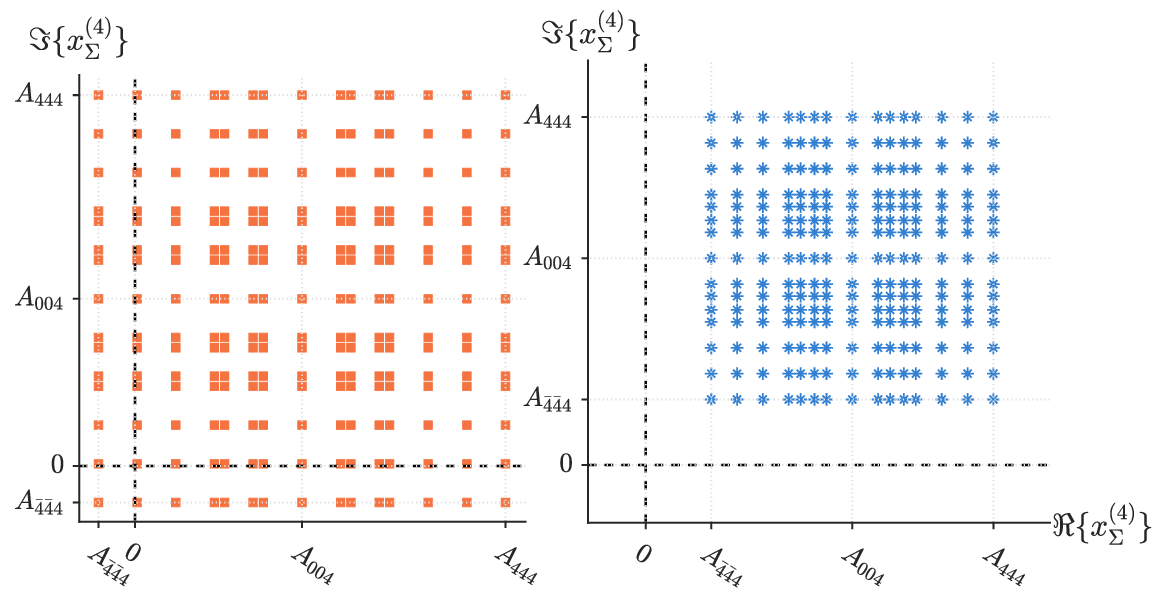}
\caption{Example of constellation diagram for  $N=3$, $\boldsymbol{T}=[4, 2, 1]$, (left) $\boldsymbol{\alpha}=[0.56, 0.32, 0.12]$, and (right) $\boldsymbol{\alpha}=[0.8,0.15,0.05]$.}
     \label{fig:Constelationinpower}
\end{figure*}

\section{PA \Rfour{and Inherent Interference Cancellation}}\label{sec:Power Optimization}
\subsection{Power Allocation}
Similar to \gls{sr}-\gls{noma}, the power allocation for \gls{mr}-\gls{noma} should prevent sub-constellation overlapping to enable reliable \gls{sic} detection \cite{iraqi2021power}. Such a conclusion is inferred from \eqref{eq:coherant Detector}, where the \gls{mr}-\gls{noma} is treated as a sequence of \gls{sr}-\gls{noma} symbols. For example, Fig.~\ref{fig:Constelationinpower} illustrates \gls{mr}-\gls{noma} constellation diagram with $N=3$,
$\boldsymbol{M}=[4,4,4]$, and 
$\boldsymbol{T}=[4, 2, 1]$. To avoid sub-constellation overlap, the
constellation point that is closest to the real and/or imaginary axis, referred to as the critical point, must not cross the $x$- or
$y$-axis \cite{iraqi2021power}. Although there are four critical points, one in each quadrant, it is sufficient to consider one of them due to the constellation symmetry. Therefore, without loss of generality, we consider the point $\{A_{\grave4\grave44}, A_{\grave4\grave44}\}$, and thus, both points must have $A_{\grave4\grave44}>0$. Using \eqref{eq:x_sigma_1} gives,
\begin{equation}
    4\sqrt{\alpha_1}-4\sqrt{\alpha_2}-4\sqrt{\alpha_3}>0
\end{equation}
 which simplifies to \cite[Eq. (22c)]{iraqi2021power},
\begin{equation}
\label{step 1}
     \sqrt{\alpha_1}> \sqrt{\alpha_{2}}  +\sqrt{\alpha_3}. 
\end{equation}

Although sub-constellation overlapping is necessary to offer reliable \gls{sic} detection, it does not guarantee certain \gls{ber} performance. Therefore, the derived \gls{ber} expressions for \gls{mr}-\gls{noma} can be used to optimize power allocation to achieve a specified \gls{ber} performance. For example, if the objective is to minimize the average \gls{ber} of all users, the problem can be formulated as,
\begin{subequations}
    \label{prob:power-min}
\begin{equation}
\boldsymbol{\alpha}^*= \arg \min_{\alpha_i\forall i\in\mathbb{R}_+^N}\quad  \bar{P}(\alpha_1,\alpha_2, \dots, \alpha_N )\tag{\ref{prob:power-min}} 
\end{equation}
\vspace{-4mm}
\begin{align}
\text{s.t.}\quad 
& \alpha_1 > \alpha_2 > \cdots > \alpha_N > 0, \label{c2:ordering}\\
& \sqrt{\alpha_1} - \sum\nolimits_{i=2}^{N}\sqrt{\alpha_i} > 0, \label{c3:separatio}\\
& \sum\nolimits_{i=1}^{N}\alpha_i = 1. \label{c4:normalization}
\end{align}
\end{subequations}
where $
\bar P=\tfrac{1}{N}\sum_{i=1}^N \bar{P}_{B_i}
$. Constraint \eqref{c2:ordering} is used to ensure that the power allocated to each user is inversely proportional
 to its average channel gain. Constraint \eqref{c3:separatio} to ensure a reliable \gls{sic}. Finally, constraint \eqref{c4:normalization} ensures that the total power is normalized to unity. The optimization problem in~\eqref{prob:power-min} is a multivariate nonlinear one with highly complex \gls{ber} expressions, and hence, it is intractable to solve it analytically. \Rone{\label{R_1_2_1}By noting that \gls{mr}-\gls{noma} is generally insensitive to deviations in power assignment from optimal, near-optimal search-based solutions can be adopted with negligible performance loss. For example, as shown in \cite[Fig. 2]{UL_MRNOMA_Arafat}, a power deviation of $\pm 0.05$ from the optimum results in \gls{ber} degradation of only $0.9\%$. Therefore, search-based methods can be considered efficient due to the limited search space and the ability to use a large search grid, denoted by $\Delta_{\alpha}$.}

 \Rfour{
 \subsection{\label{R_1_3_2}Inherent Interference Cancellation}
\label{Inherent Interferacne}
Due to its structure, \gls{mr}-\gls{noma} inherently performs interference cancellation, which may significantly improve its performance. To clarify, the \gls{sinr} analysis is presented for the $N=2$ case with \gls{bpsk}, and $\boldsymbol{T}=[2,1]$. Based on the coherent-combining detector model in \eqref{eq:coherant Detector}, the \gls{sinr} depends on the transmitted signal pattern of $U_2$. For the considered configuration, there are only two possible patterns: $s_{2,1}[1]=s_{2,2}[2]$ and $s_{2,1}[1]\neq s_{2,2}[2]$. Therefore, for the equal symbols case, it is straightforward to show that
\begin{equation}
\zeta=\frac{|h_{1}|^{2}\alpha
_{1}}{|h_{1}|^{2}\alpha _{2}+0.5\sigma_{\mathrm{w}}^{2}}, \quad s_{2,1}[1]=s_{2,2}[2]. 
\end{equation}
For the second scenario of unequal symbols' values, 
\begin{equation}
\zeta=\frac{2|h_{1}|^{2}\alpha_{1}}{\sigma
_{\mathrm{w}}^{2}}, \quad s_{2,1}[1] =- s_{2,2}[2]
\end{equation}
which implies that the interference is eliminated. 
Averaging over both equiprobable cases, 
\begin{equation}
\zeta  =\frac{|h_{1}|^{2}\alpha _{1}}{2|h_{1}|^{2}\alpha _{2}+\sigma _{\mathrm{w}%
}^{2}}+\frac{|h_{1}|^{2}\alpha _{1}}{\sigma _{\mathrm{w}}^{2}}\triangleq \zeta_1+\zeta_2.
\label{eq:sinr_avg}
\end{equation}%
Thus, \gls{sinr} for \gls{mr}-\gls{noma} consists of two parts, $\zeta_1$ is similar to the \gls{sinr} of \gls{sr}-\gls{noma}, except that the interference term is scaled by a factor of $2$ while $\zeta_2$ is equivalent to the \gls{snr} of \gls{oma}. At high \glspl{snr} $\zeta_2\gg \zeta_1$ and thus the \gls{mr}-\gls{noma} has much higher \gls{sinr} compared to \gls{sr}-\gls{noma}, i.e. $\lim_{\bar{\gamma}\rightarrow \infty} \zeta(\text{MR-NOMA})/\zeta(\text{SR-NOMA})=\infty$, where the average \gls{snr} is defined as $\bar{\gamma}=1/\sigma^{2}_{\mathrm{w}}$.
At low \glspl{snr}, the difference becomes smaller because the performance is dominated by \gls{awgn}, therefore, $\lim_{\bar{\gamma}\rightarrow  0} \zeta(\text{MR-NOMA})/\zeta(\text{SR-NOMA})=2$, which is $3$ dB. Hence, the inherent interference cancellation in \gls{mr}-\gls{noma} gives a significant advantage that \gls{sr}-\gls{noma} does not have.} 
\begin{figure*}
    \centering
\includegraphics[width=0.95\linewidth]{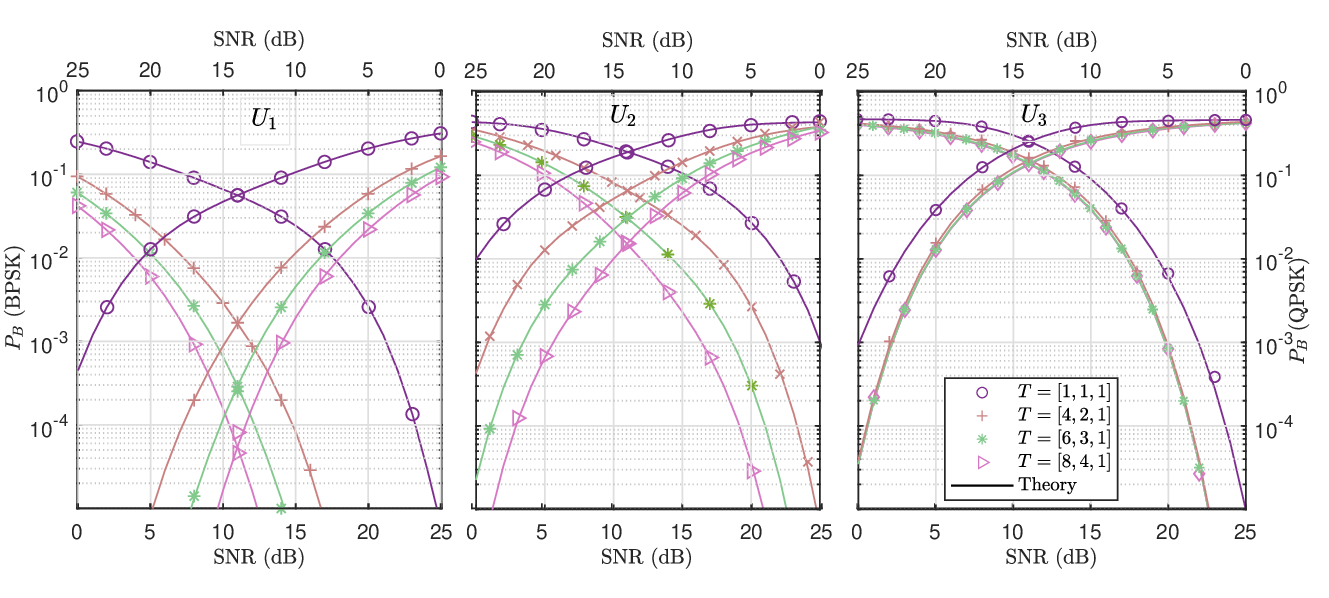}
\vspace{-0.5em}
    \caption{Exact conditional \gls{ber} with various symbol periods for \gls{bpsk} (left y-axis)
and \gls{qpsk} (right y-axis).}
    \label{fig:CondetionalBERQPSK}
     \vspace{-0.5em}
\end{figure*}

\section{Numerical Results} \label{sec:Numerical Results}

\begin{table}[t] 
\color{RTwoColor}
\centering 
\caption{\Rtwo{Default simulation parameters.}} \label{tab:simulation_parameters} 
\renewcommand{\arraystretch}{0.95}
\begin{tabular}{l c} \hline \textbf{Parameter} & \textbf{Value} \\ \hline Number of users, $N$ & $3$ \\ Modulation-order vector, $\boldsymbol{M}$ & $[4,4,4]$ \\ Power-coefficient vector, $\boldsymbol{\alpha}$ & $[0.8,0.15,0.05]$ \\ User-distance vector, $\boldsymbol{d}$ & $[3,2,1]$ \\ Nakagami fading parameter, $\tilde{m}$ & $2$ \\ Path-loss exponent, $\eta$ & $2$ \\ Maximum transmit power, $P_t$ & $1$ \\ Minimum symbol duration, $T_N$ & $1$ \\ Monte Carlo realizations & $10^{6}$ \\ \hline \end{tabular} 
\vspace{-1em}
\end{table}
The downlink performance of the proposed \gls{mr}-\gls{noma} scheme is evaluated over a Nakagami-$\frak{m}$ fading channel, and  $\sigma^{2}_{\mathrm{w}}$ is assumed identical for all users. 
\Rtwo{ \label{R_2_6}Unless stated otherwise, the simulation parameters are given in Table \ref{tab:simulation_parameters}. In the figures, markers denote the simulation results, whereas solid and dashed lines represent the analytical results, \Rfour{\label{R_4_7} also common curve labels are shown once in the shared legend, while subplot-specific labels are indicated inside each subplot}. Each simulation point is averaged over $10^{6}$ Monte Carlo trials. All Monte Carlo simulation results are obtained using imperfect \gls{sic}, while the analytical curves follow the \gls{sic} assumption stated in the related figure discussion. \Rtwo{ For more details of the simulation setup, the MATLAB codes for the analytical and simulation results are available in \cite{Paper_Code}}}. \Rfour{\Rtwo{}\label{R_4_8}The performance of \gls{mr}-\gls{noma} is compared with the downlink \gls{sr}-\gls{noma}, which is labeled as $\boldsymbol{T}=[1,\dots 1]$}. 

Fig.~\ref{fig:CondetionalBERQPSK} presents the conditional \gls{ber} versus \gls{snr} for the imperfect \gls{sic} case with $N=3$ using 
\gls{bpsk} and \gls{qpsk}. The left and right $y$-axis represent the \gls{ber} of \gls{bpsk} and \gls{qpsk}, respectively. As the figure shows, the simulation and analytical results agree perfectly, validating the accuracy of the derived \gls{ber} expressions for both modulation schemes. 
The results further reveal a noticeable improvement in the \gls{ber} of $U_1$ as the symbol rate decreases, which can be attributed to the higher symbol energy associated with longer symbol intervals.  \Rone{\label{R_1_3_3}Further, the figure shows a significant \gls{ber} difference between \gls{mr} and \gls{sr}-\gls{noma}. For example, at $P_B=10^{-5}$, the \gls{snr} gains for $U_1$ are approximately $12.3~\mathrm{dB}$ and $10.4~\mathrm{dB}$ for $\mathbf{T}=[8,4,1]$ and $\mathbf{T}=[6,3,1]$, respectively. These gains are mainly due to the symbol-energy advantages $10\log_{10}(8)=9.03~\mathrm{dB}$ and $10\log_{10}(6)=7.78~\mathrm{dB}$, together with interference averaging. Specifically, using \eqref{eq:sinr_avg} with $\boldsymbol{\alpha}=[0.8,0.15,0.05]$, $|h_1|^2=1$, and an input \gls{snr} of $15~\mathrm{dB}$ gives $\zeta_{[1,1,1]}=5.38~\mathrm{dB}$, $\zeta_{[8,4,1]}=9.73~\mathrm{dB}$, and $\zeta_{[6,3,1]}=9.56~\mathrm{dB}$, corresponding to \gls{sinr} improvements of $4.34~\mathrm{dB}$ and $4.17~\mathrm{dB}$, respectively. A similar trend is observed for $U_2$, whereas the improvement for $U_3$ is limited because $T_3=1$ in all cases and mainly results from more reliable \gls{sic} of $U_1$ and $U_2$, which nearly converge to the perfect \gls{sic} bound, that represents the optimum interference cancellation case.}  
It is worth mentioning that the results in Fig.~\ref{fig:CondetionalBERQPSK} indicate that increasing the modulation order requires higher \gls{snr} values to achieve a fixed \gls{ber} threshold. For instance, with $\boldsymbol T=[4,2,1]$, a target \gls{ber} of $10^{-2}$ is attained by $U_1$ at approximately 7~dB for \gls{bpsk}, whereas \gls{qpsk} requires around 10~dB to achieve the same performance.

\begin{figure}[t]
    \centering
\includegraphics[width=0.95\linewidth]{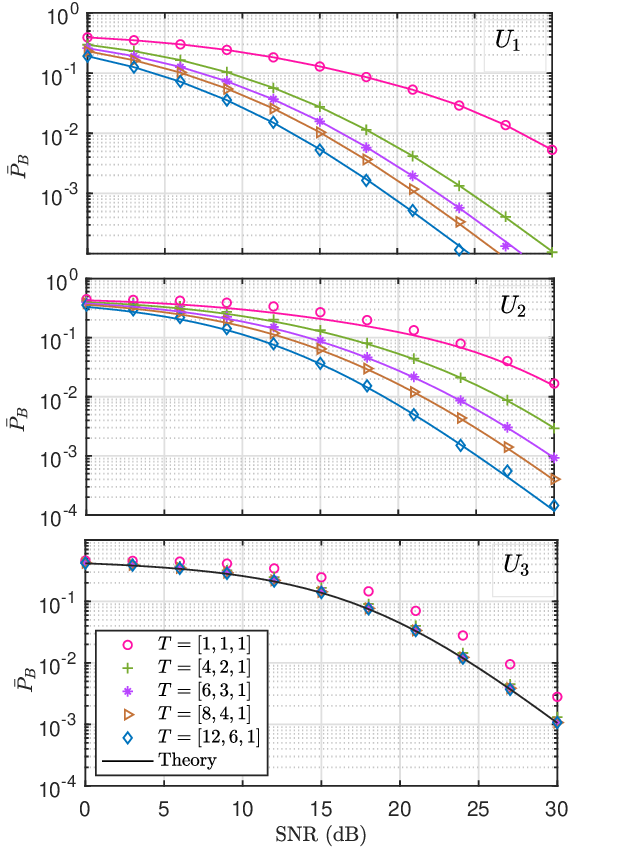}
\vspace{-0.9em}
    \caption{Average \gls{ber} using the perfect \gls{sic} approximation.}  \label{fig:AverageBER 3users}
     \vspace{-0.7em}
\end{figure}

Fig.~\ref{fig:AverageBER 3users} illustrates the average \gls{ber} performance over a Nakagami-$\frak{m}$ fading channel with $\frak{m}=2$, where \gls{qpsk} modulation is employed for $U_1$, $U_2$, and $U_3$, respectively. The theoretical curves are obtained under the assumption of perfect \gls{sic} for $U_2$ and $U_3$. Specifically, the results of $U_1$ show a clear processing gain as the symbol duration increases, owing to the corresponding increase in symbol energy. \Rone{Further, the gap between the imperfect-\gls{sic} simulation and the perfect-\gls{sic} analytical result of $U_2$ and $U_3$ is smaller in the considered \gls{mr}-\gls{noma} cases, whereas this gap is more pronounced in conventional \gls{sr}-\gls{noma}, where the additional symbol-energy gain is not available. Such performance is obtained because as the symbol duration increases, the detected symbols benefit from higher accumulated symbol energy, which improves the reliability of the \gls{sic} process}. However, the relative improvement for $U_3$ among different \gls{mr}-\gls{noma} configurations is limited because $T_3=1$ in all cases, which makes the performance of \gls{mr}-\gls{noma} to be almost identical across configurations and nearly converge to the perfect \gls{sic} bound, which represents the optimum interference cancellation case. Specifically, at $P_b \approx 2.8\times10^{-2}$, the required \glspl{snr} are $21.5$ and $24\,\mathrm{dB}$ for \gls{mr}-\gls{noma} and \gls{sr}-\gls{noma}, respectively, with at most $0.3\,\mathrm{dB}$ spread across $\boldsymbol{T}=[4,2,1],[6,3,1],[8,4,1],[12,6,1]$.

\begin{figure}
    \centering
\includegraphics[width=0.95\linewidth]{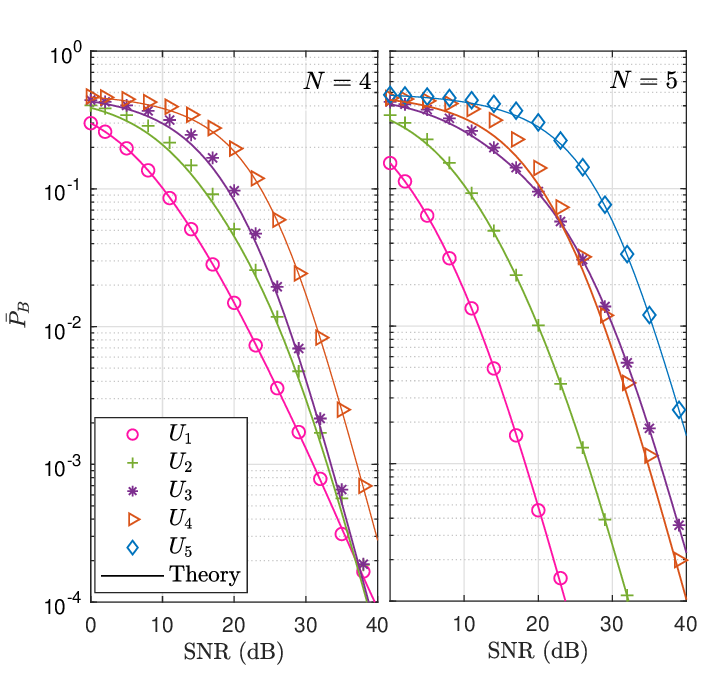}
    \caption{Average \gls{ber} using \eqref{eq:U_1 Average BER for arbitrary symbol duration and number of users}, $M_i=4~\forall i$, for $N=4$, $\boldsymbol T=[8,4,2,1]$, $\boldsymbol{\alpha}=[0.7, 0.22, 0.07, 0.01]$, $\mathbf{d}=[4, 3, 2, 1]$,  For $N=5$, $\boldsymbol T=[8,4,2, 2,1]$, $\boldsymbol{\alpha}=[0.8, 0.16, 0.03, 0.009, 10^{-3}]$, $\mathbf{d}=[2, 1.5, 1, 0.75, 0.5]$.}  \label{fig:AverageBER 4users}
     \vspace{-1.5em}
\end{figure}

\Rone{Fig.~\ref{fig:AverageBER 4users} extends the study to a four- and five-user \gls{mr}-\gls{noma} system, where all users employ \gls{qpsk} modulation. As shown in Fig.~\ref{fig:AverageBER 4users}, the simulation results align closely with the analytical curves obtained under the perfect \gls{sic} assumption. Moreover, for $N=4$ compared with the three-user case, adding a fourth user naturally increases the interference level. More specifically, although we assume a perfect \gls{sic} in the analytical expression, the analytical and simulation results shown remain in close agreement, with the accuracy of the match improving as the symbol energy increases. Further, it is observed that the agreement for $U_2$ is tighter than for $U_3$, which can be attributed to the higher symbol energy assigned to $U_2$ across all studied cases. Moreover, as the symbol energy grows, the analytical and simulation curves converge even further. While for $U_4$, the \gls{ber} curves across all symbol duration settings are nearly the same and exhibit close agreement with the perfect \gls{sic} curve. For $N=5$, a similarly close agreement is observed between the analytical and simulation results for all users.}
\begin{table}[t]
\centering
\setlength{\tabcolsep}{4pt}
\renewcommand{\arraystretch}{1.15}
\scriptsize
\caption{Optimal power coefficients for various SNRs and symbol-duration vectors, with $\Delta_{\alpha}=10^{-3}$. The results are rounded to three decimal places.}
\label{tab:alpha_min}
\begin{tabular}{|c|c|c|c|c|c|}
\hline
\textbf{\gls{snr} (dB)}
& $\boldsymbol{T}$
& $\boldsymbol{\alpha_{1}^{*}}$
& $\boldsymbol{\alpha_{2}^{*}}$
& $\boldsymbol{\alpha_{3}^{*}}$
& $\boldsymbol{\bar{P}^{\star}}$ \\
\hline

\multirow{3}{*}{10}
& $[1,1,1]$ & 0.656 & 0.245 & 0.099 & 0.180 \\ \cline{2-6}
& $[4,2,1]$ & 0.660 & 0.228 & 0.112 & 0.108 \\ \cline{2-6}
& $[8,4,1]$ & 0.664 & 0.205 & 0.131 & 0.072 \\
\hline
\multirow{3}{*}{15}
& $[1,1,1]$ & 0.712 & 0.234 & 0.054 & 0.100 \\ \cline{2-6}
& $[4,2,1]$ & 0.659 & 0.258 & 0.083 & 0.042 \\ \cline{2-6}
& $[8,4,1]$ & 0.660 & 0.228 & 0.112 & 0.022 \\
\hline

\multirow{3}{*}{20}
& $[1,1,1]$ & 0.780 & 0.188 & 0.032 & 0.037 \\ \cline{2-6}
& $[4,2,1]$ & 0.706 & 0.234 & 0.060 & 0.011 \\ \cline{2-6}
& $[8,4,1]$ & 0.685 & 0.223 & 0.092 & 0.004 \\
\hline

\multirow{3}{*}{25}
& $[1,1,1]$ & 0.799 & 0.175 & 0.026 & 0.008 \\ \cline{2-6}
& $[4,2,1]$ & 0.725 & 0.223 & 0.052 & 0.002 \\ \cline{2-6}
& $[8,4,1]$ & 0.701 & 0.217 & 0.082 & 0.001 \\
\hline
\end{tabular}
\vspace{-0.9em}
\end{table}

\begin{figure}
    \centering
\includegraphics[width=0.9\linewidth]{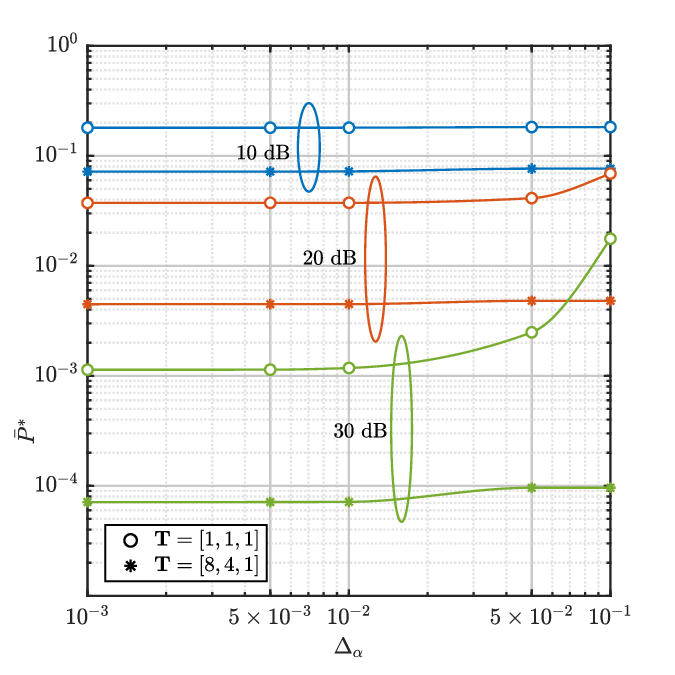}
    \caption{Impact of the search step $\Delta_{\alpha}$ on $\bar{P}^*$.}
    \label{fig:BiS vs EGS}
    \vspace{-1.5em}
\end{figure}

\Rone{
Fig.~\ref{fig:BiS vs EGS} presents the optimized average \gls{ber},
$\bar{P}^{\star}$, obtained using \gls{egs}-\gls{pa} as a function of
$\Delta_{\alpha}$ for different \gls{snr} values. Across all SNRs,
$\bar{P}^{\star}$ remains nearly unchanged at fine resolutions but increases
as $\Delta_{\alpha}$ approaches $10^{-1}$, with the most pronounced
degradation observed at $30\,\mathrm{dB}$. Specifically, at
$\Delta_{\alpha}=10^{-1}$, the curve for
$\boldsymbol{T}=[1,1,1]$ rises more steeply than that for
$\boldsymbol{T}=[8,4,1]$, because the larger entries of
$\boldsymbol{T}$ in \gls{mr}-\gls{noma} cause the remaining feasible
power-allocation solutions to produce nearly similar \gls{ber} values that remain
close to the optimum, despite the coarse grid reducing the number of
feasible points in both \gls{sr}- and \gls{mr}-\gls{noma}.}

\begin{figure}
     \centering
 \includegraphics[width=0.85\linewidth]{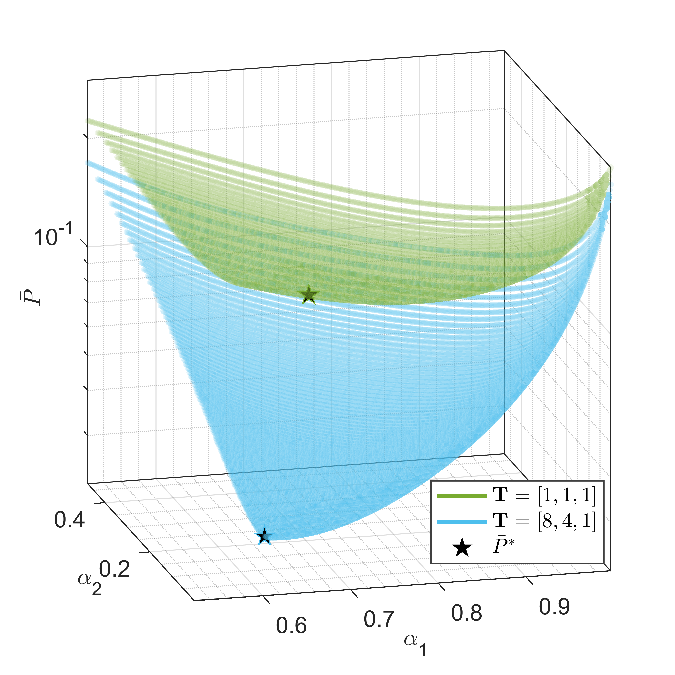} \caption{$\bar{P}$ vs $\alpha_1$ and $\alpha_2$, \gls{snr}$=15$ dB, and $\Delta_\alpha=10^{-3}$.}
     \label{fig:3D Power Optimization}
     \vspace{-1.5em}
 \end{figure}

Fig. \ref{fig:3D Power Optimization} depicts the surface $\bar{P}(\alpha_1,\alpha_2)$ at $\mathrm{SNR}=15$\, dB for the three–user case under two symbol–duration configurations, $\boldsymbol T=[1,1,1]$ and $[8,4,1]$, using $\Delta_\alpha=10^{-3}$, the feasible set is determined by the power coefficients constraints in \eqref{c2:ordering}, and \eqref{c3:separatio}, and the total power normalization constraint  \eqref{c4:normalization}. The star markers highlight the minimum values. Relative to the \gls{sr} case, the \gls{mr} surface exhibits a deeper valley and a uniformly lower plateau, indicating that symbol energy in \gls{mr}-\gls{noma}  reduces the average error probability. Moreover, Table~\ref{tab:alpha_min} breaks down the optimal power coefficients, $\boldsymbol \alpha^*$, and $\bar{P}^\star$, versus various \gls{snr} values and symbol duration sets. For every configuration, the minimum average \gls{ber} $\bar{P}^{\star}$ decreases monotonically with \gls{snr}. Moreover, at a fixed \gls{snr}, moving from \gls{sr} to \gls{mr}  yields smaller $\bar{P}^{\star}$, for example, at $15$\,dB, $\bar{P}^{\star}$ decreases from $0.1$ at $\boldsymbol T=[1,1,1]$ to $0.022$ at $\boldsymbol T=[8, 4, 1]$.

\section{conclusion {and Future Work}}\label{section:Conlusion}
This work investigated the downlink performance of \gls{mr}-\gls{noma} over Nakagami-$\frak{m}$ fading channels for arbitrary symbol durations and numbers of users. An exact \gls{ber} expression was derived for the first user, irrespective of the total number of users. For the three-user case, accurate closed-form BER approximations were derived for the second and third users while accounting for imperfect \gls{sic}. For the generalized case with more than three users, however, the mathematical complexity associated with imperfect \gls{sic} renders the analysis intractable, therefore, the corresponding \gls{ber} expressions were derived under the assumption of perfect \gls{sic}. The optimum power allocation problem was also formulated as a nonlinear optimization problem and solved using a linear search method. 
The obtained results demonstrated that \gls{mr}-\gls{noma} provides greater flexibility in user pairing while maintaining reliable \gls{ber} performance. This flexibility constitutes an important advantage over \gls{sr}-\gls{noma}, in which all users are constrained to operate at the same symbol rate. 

The \gls{ber} expressions presented in this work were derived for \gls{bpsk} and \gls{qpsk} modulation schemes. { Therefore, extending the analysis to scenarios in which users employ different modulation orders and symbol rates represents a promising direction for future research. Unlike conventional \gls{noma}, the error-rate analysis of generalized \gls{mr}-\gls{noma} with higher-order modulation requires accounting for a prohibitively large number of symbol and interference combinations, making a direct analytical approach impractical. Therefore, developing a tractable and scalable framework for analyzing the error-rate performance of generalized \gls{mr}-\gls{noma} systems with higher-order modulation remains an important open research problem.} \Rtwo{The concept of \gls{mr} can be applied to other non-orthogonal transmission schemes such as \gls{rsma}. The natural way to apply the \gls{mr} concept to \gls{rsma} is to introduce heterogeneous symbol rates into the common/private-stream architecture of \gls{rsma}. This would create what could be termed \gls{mr}-\gls{rsma}.}

\bibliography{mRef3}
\bibliographystyle{IEEEtran}
\appendices
\section{}
\label{app:symbols}
\small \Rtwo{


\nomenclature{\(N\)}{Total number of users}
\nomenclature{\(U_n\)}{The \(n\)th user, \(n\in\{1,2,\ldots,N\}\)}
\nomenclature{\(\boldsymbol{T}\)}{Symbol-duration vector, \(\boldsymbol{T}=[T_1,T_2,\ldots,T_N]\)}


\nomenclature{\(\delta_n\)}{Number of symbols for \(U_n\) in \(T_1\) interval, \(\delta_n=T_1/T_n\)}
\nomenclature{\(\ell\)}{Signaling-period or sample index}

\nomenclature{\(\mathbf{S}_N\)}{Data-symbol matrix of the \(N\) users over \(T_1\)}
\nomenclature{\(\mathbf{s}\)}{Data-symbol vector in \(T_1\)}

\nomenclature{\(\mathbf{x}\)}{Transmitted MR-NOMA vector}
\nomenclature{\(\mathbf{x}_n\)}{Contribution of \(U_n\) to the transmitted MR-NOMA vector}
\nomenclature{\(P_t\)}{Maximum transmit power}

\nomenclature{\(\mathbf{r}\)}{Received signal vector, $\mathbf{r}=[r[1],r[2],\dots,r[N]]$}
\nomenclature{\(h_n\)}{Channel fading coefficient}
\nomenclature{\(\mathbf{d}\)}{User-distance vector, \(\mathbf{d}=[d_1,d_2,\ldots,d_N]\)}
\nomenclature{\(\eta\)}{Path-loss exponent}
\nomenclature{\(\mathbf{w}\)}{AWGN vector, $\mathbf{w}_n=[w_n[1], w_n[2],\dots, w_n[N]]$}
\nomenclature{\(\sigma_w^2\)}{AWGN noise power}

\nomenclature{\(\hat{s}\)}{Estimated data symbol}
\nomenclature{\(\acute{r}\)}{Received vector after SIC processing}

\nomenclature{\(P_{B}\)}{Conditional BER}
\nomenclature{\(C_j\)}{The \(j\)th possible transmitted-symbol case}
\nomenclature{\(\grave{a}\)}{Negative value of \(a\), i.e., \(\grave{a}\triangleq-a\)}
\nomenclature{\(x_{\Sigma}^{(v)}\)}{Sum of \(v\) transmitted samples}
\nomenclature{\(w^{(V)}\)}{Sum of \(V\) AWGN samples}

\nomenclature{\(\gamma\),$\bar{\gamma}$}{Instantaneous and average \gls{snr}} 
\nomenclature{\(G_{m,n}\)}{Detection scaling factor, defined as \(\sqrt{\gamma_m/T_n}\)}
\nomenclature{\(\boldsymbol{M}\)}{Modulation-order vector, $\boldsymbol{M}=[M_1, M_2,\dots,M_N]$}
\nomenclature{\(\kappa_n\)}{Modulation-dependent normalization factor}
\nomenclature{\(E^{(T_n,T_m)}\)}{Ordered set \(=[-T_m,-T_m+2T_n,\ldots,T_m]\)}
\nomenclature{\(I^{(T_n)}\)}{Ordered set \(=[-T_n,-T_n+2,\ldots,T_n]\)}

\nomenclature{\(\nabla\)}{SIC outcome indicator}

\nomenclature{\(\lambda\)}{Channel gain, \(\lambda=\lvert h\rvert^2\)}
\nomenclature{\(\frak{m}\)}{Nakagami fading parameter}
\nomenclature{\(\Omega\)}{Average gain of the channel}
\nomenclature{\(\mathbb{E}[\cdot]\)}{Expectation operator}
\nomenclature{\(\Gamma(\cdot)\)}{Gamma function}
\nomenclature{\(\bar{P}_{B}\)}{Average BER}

\nomenclature{\(\bar{P}\)}{Average BER of all users}
\nomenclature{\(\rho\)}{BER threshold}
\nomenclature{\(\alpha^\ast\)}{Optimal power coefficient}
\nomenclature{\(P^\star\)}{Minimum average BER}

\printnomenclature
}

\section{BER Expressions for User~3}
\label{app:U3-BER}
\begin{table}[t]
\centering
\small
\caption{Parameters for \eqref{eq:user1correct,user2correct_symbols(a,z,q)}--\eqref{eq:user1wrong,user2wrong_symbols(a,b,c)}}
\label{tab:U3-params-azg}
\resizebox{0.35\textwidth}{!}{%
\begin{tabular}{|c||c|c|c|c|c|c|c|c|c|}
\hline
$j$ & $a_{j}$ & $z_{j}^{(1)}$ & $z_{j}^{(2)}$ & $z_{j}^{(3)}$ & $g_{j}^{(1)}$ & $g_{j}^{(2)}$ &  $u_{j}^{(1)}$ & $u_{j}^{(2)}$\\
\hline\hline
1  & $A_{\grave{4}\grave{4}4}$ & 0 & 0 & 2 & 0 & 1 & 1 & 0 \\ \hline
2  & $A_{\grave{2}\grave{4}4}$ & 0 & 1 & 1 & 1 & 3 & 3 & 1 \\ \hline
3  & $A_{0\grave{4}4}$         & 1 & 2 & 1 & 3 & 3 & 3 & 3 \\ \hline
4  & $A_{2\grave{4}4}$         & 1 & 1 & 0 & 3 & 1 &  1 & 3 \\ \hline
5  & $A_{4\grave{4}4}$         & 2 & 0 & 0 & 1 & 0 & 0 & 1 \\ \hline
6  & $A_{\grave{4}44}$         & 2 & 0 & 0 & 1 & 0 &  1 & 0 \\ \hline
7  & $A_{\grave{2}44}$         & 1 & 1 & 0 & 3 & 1 & 3 & 1 \\ \hline
8  & $A_{044}$                 & 1 & 2 & 1 & 3 & 3 &  3 & 3 \\ \hline
9  & $A_{244}$                 & 0 & 1 & 1 & 1 & 3 &  1 & 3 \\ \hline
10 & $A_{444}$                 & 0 & 0 & 2 & 0 & 1 &  0 & 1\\ \hline
11 & $A_{\grave{4}04}$         & 1 & 0 & 1 & 1 & 1 & 1 & 0 \\ \hline
12 & $A_{\grave{2}04}$         & 1 & 2 & 1 & 4 & 4 &  3 & 2 \\ \hline
13 & $A_{004}$                 & 2 & 2 & 2 & 6 & 6 &  3 & 6 \\ \hline
14 & $A_{204}$                 & 1 & 2 & 1 & 4 & 4 & 1 & 6 \\ \hline
15 & $A_{404}$                 & 1 & 1 & 0 & 1 & 1 &  0 & 2 \\ \hline
\end{tabular}
}
\end{table}

%
\begin{table}[!t]
\centering
\small
\caption{Parameters for \eqref{eq:user1wrong,user2correct_symbol} and \eqref{eq:user1wrong,user2wrong_symbols(a,b,c)}.}
\label{tab:U3-params-bu}
\resizebox{0.35\textwidth}{!}{%
\begin{tabular}{|c||c|c|c|c|c|c|c|c|c|c|}
\hline
$j$ & $b^{(1)}_{j}$ & $b^{(2)}_{j}$ & $b^{(3)}_{j}$ & $b^{(4)}_{j}$ & $b^{(5)}_{j}$ & $b^{(6)}_{j}$ & $c^{(1)}_{j}$ & $c^{(2)}_{j}$ & $c^{(3)}_{j}$ & $c^{(4)}_{j}$ \\ \hline
\hline
1  & 2 & 0 & 0 & 0 & 0 & 0 & 1 & 0 & 0 & 0 \\ \hline
2  & 1 & 1 & 0 & 0 & 0 & 0 & 3 & 1 & 0 & 0 \\ \hline
3  & 1 & 2 & 1 & 0 & 0 & 0 & 3 & 3 & 0 & 0 \\ \hline
4  & 0 & 1 & 1 & 0 & 0 & 0 & 1 & 3 & 0 & 0 \\ \hline
5  & 0 & 0 & 2 & 0 & 0 & 0 & 0 & 1  & 0 & 0\\ \hline
6  & 0 & 0 & 0 & 2 & 0 & 0 & 0 & 0 & 1 & 0 \\ \hline
7  & 0 & 0 & 0 & 1 & 1 & 0 & 0 & 0 & 3 & 1 \\ \hline
8  & 0 & 0 & 0 & 1 & 2 & 1 & 0 & 0 & 3 & 3 \\ \hline
9  & 0 & 0 & 0 & 0 & 1 & 1 & 0 & 0 & 1 & 3 \\ \hline
10 & 0 & 0 & 0 & 0 & 0 & 2 & 0 & 0 & 0 & 1 \\ \hline
11 & 1 & 0 & 0 & 1 & 0 & 0 & 2 & 0 & 1 & 0 \\ \hline
12 & 1 & 1 & 0 & 1 & 1 & 0 & 6 & 1 & 3 & 1 \\ \hline
13 & 1 & 1 & 1 & 1 & 1 & 1 & 6 & 3 & 3 & 3 \\ \hline
14 & 0 & 1 & 1 & 0 & 1 & 1 & 2 & 3 & 1 & 3 \\ \hline
15 & 0 & 0 & 1 & 0 & 0 & 1 & 0 & 1 & 0 & 1 \\ \hline
\end{tabular}
}
\end{table}

In this appendix, we present the expressions of the four
conditional terms that compose the \gls{ber} of $U_3$ in \eqref{eq:general:U3 BER}, as follows,
the case where both $U_1$ and $U_2$ are correctly detected is given by,
\begin{multline}
\sum_j\Pr(\mathrm{E}_3|[\nabla^1_S,\nabla^2_S,\mathcal{C}_j])
=
\sum_j4\Big[F\left(a_jG_{3,1}\right) \\
\times \sum_{i=1}^3  z_j^{(i)}
      F\left(A_{E^1_{i+1}20}G_{3,2}\right)
      Q\left(A_{100}G_{3,3}\right)\Big].
\label{eq:user1correct,user2correct_symbols(a,z,q)}
\end{multline}

When $U_1$ is correctly detected and $U_2$ is erroneously
detected is,
\begin{multline}
\sum_j\Pr(\mathrm{E}_3|[\nabla^2_F,\nabla^1_S,\mathcal{C}_j])
=
\sum_j\Big[F\!\left(a_j G_{3,1}\right) \\
\times \sum_{i=1}^3 z_j^{(i)}
      Q\left(A_{E^1_{i+1}24}G_{3,2}\right)
      \big(g_j^{(1)}F(A_{\grave120}G_{3,3}) \\
      + g_j^{(2)}Q(A_{120}G_{3,3})\big)\Big].
\label{eq:user1correct,user2wrong_symbols(a,z,g)}
\end{multline}

When $U_1$ is erroneously detected and $U_2$ is correctly
detected is,
\begin{multline}
\label{eq:user1wrong,user2correct_symbol}
\sum_j\Pr(\mathrm{E}_3|[\nabla^2_S,\nabla^1_F,\mathcal{C}_j])
=
\sum_j\bigg[Q(a_jG_{3,1}) \\
\times\Big(\sum_{i=1}^3 b^{(i)}_j
              Q\left(A_{E^1_{i+1}\grave24}G_{3,2}\right)
          + b^{(i+3)}_j
              F\left(A_{E^1_{i+1}24}G_{3,2}\right)\Big) \\
\times \Big(u_j^{(1)}F(A_{\grave102}G_{3,3})
          + u_j^{(2)}Q(A_{102}G_{3,3})\Big)\bigg].
\end{multline}

Finally, the case where both $U_1$ and $U_2$ are erroneously detected is
\begin{multline}
\label{eq:user1wrong,user2wrong_symbols(a,b,c)}
\sum_j\Pr(\mathrm{E}_3|[\nabla^2_F,\nabla^1_F,\mathcal{C}_j])
=
\sum_j\Big[Q(a_{j}G_{3,1}) \bigg(\sum_{i=1}^3 b^{(i)}_j \\
\times F\big(A_{E^1_{i+1}\grave24}G_{3,2}\big)
+ b^{(i+3)}_jQ\left(A_{E^1_{i+1}24}G_{3,2}\right)\bigg) \\
\times\Big(c_j^{(1)}F(A_{\grave1\grave22}G_{3,3})
+ c_j^{(2)}Q(A_{1\grave22}G_{3,3})
+ c_j^{(3)}F(A_{\grave122}G_{3,3}) \\
+ c_j^{(4)}Q(A_{122}G_{3,3})\Big)\Big].
\end{multline}
The parameters $a_j$, $z_j^{(i)}$, $g_j^{(i)}$, $b_j^{(i)}$,
$u_j^{(i)}$, and $c_j^{(i)}$ used in
\eqref{eq:user1correct,user2correct_symbols(a,z,q)}–%
\eqref{eq:user1wrong,user2wrong_symbols(a,b,c)}
are listed in Tables~\ref{tab:U3-params-azg} and~\ref{tab:U3-params-bu}. Interested readers
can refer to the MATLAB code \cite{Paper_Code} to compute the BER
expressions.


\end{document}